\documentclass[a4paper,twocolumn]{spaceDebrisC} 
\usepackage{times}
\usepackage[numbers]{natbib}
\usepackage{graphicx}
\usepackage{subcaption}
\usepackage{amsmath}
\usepackage{amssymb}
\usepackage{xcolor}

\title{A tri-static ground-based laser ranging method for precise satellite attitude determination}
\author{Peter Bartram}
\author{Jim Fletcher}
\author{Ewan Schafer}
\author{Liam Pieters}
\author{David Gooding}
\author{Hira Virdee}
\affil{Lumi Space, United Kingdom, Emails: \{pete, jim, ewan, liam, dave, hira\}$@$lumi.space}

\begin{document}


\maketitle

\begin{abstract}
The ability to accurately determine the rotation rate and spin axis of active satellites during deployment, during phases of uncertain operations (e.g. loss of control, potential fuel leaks) and particularly for defunct satellites to assess suitability of removal of space objects by active debris removal is increasingly important. Conventional techniques, such as optical photometry, face challenges in providing precise attitude information. In this work, we propose a tri-static ground-based satellite laser ranging (SLR) approach combined with three onboard laser retroreflectors for precise satellite attitude determination. Via this approach our method achieves true 3D triangulation of the reflectors’ positions in space, enabling a highly accurate estimation of the satellite’s attitude. This multi-station configuration overcomes limitations of existing single-station SLR techniques that rely on indirect inference, e.g. Fourier analysis of range residuals, or lead to multiple possible attitude solutions. We demonstrate through simulations that our approach can estimate the  spin rate and axis of a satellite with high precision, on the order of 0.1°/s  and 1°, respectively. Notably, this can be done with data from only a single pass and does not require observation of a whole satellite rotation period. We show how the layout of retroreflectors are paramount to the effectiveness of the method. We provide placement strategies to maximize attitude determination performance to allow operators to incorporate retroreflectors effectively into their own satellites. These contributions lay the groundwork for a complete ground-based solution for attitude determination, significantly improving current methods and directly supporting ADR efforts.

\end{abstract}
\section{Introduction}

\begin{figure}
\centering
\includegraphics[width=\linewidth]{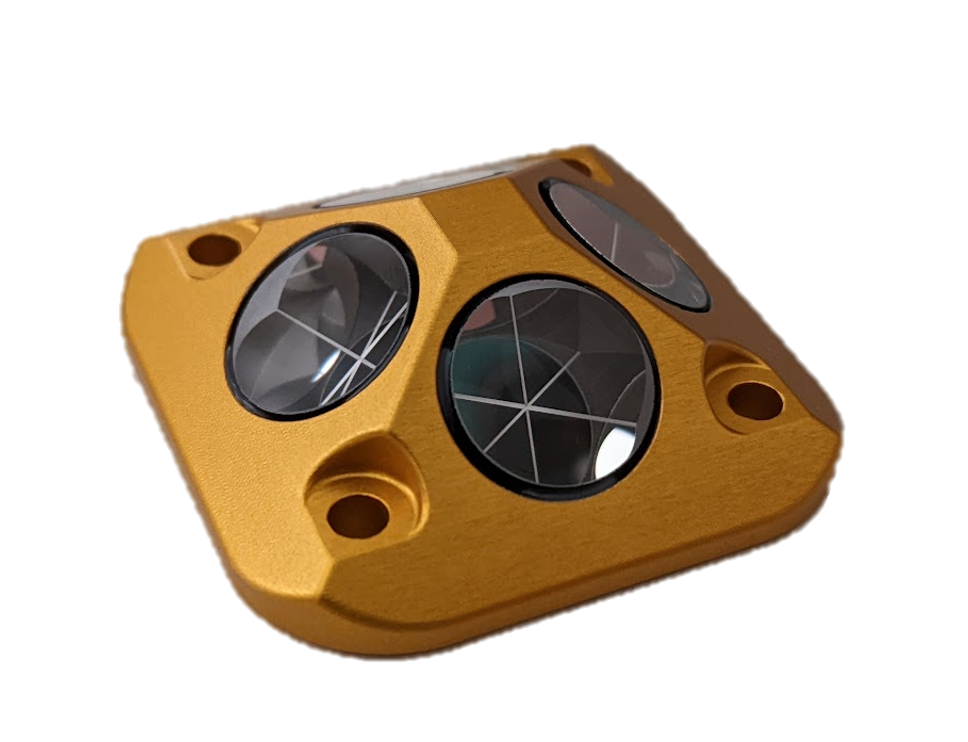}
\caption{The Lumi Micro-1 commercial off-the-shelf retroreflector. These weigh $20$~g and can be placed on the body of a satellite to allow for precision tracking even for defunct satellites. The placement of three retroreflectors onboard can be used with the methods in this paper to obtain highly precise attitude estimates.}
\label{img: gold retro}
\end{figure}

Preventing cascading collision of objects in orbit and ensuring the sustainability of the LEO space environment already necessitates the removal of $5$-$10$ pieces of debris per year via active debris removal (ADR). ADR missions use chaser satellites to match the orbits of defunct space objects and then use one of many possible techniques \cite{Poozhiyil2023} to capture the target object before bringing it into the upper atmosphere for reentry. One of the key challenges in capturing an uncooperative object is accurately assessing the rotation of the target \cite{Howlett2023}. Spin rates that are too high can lead to mechanical failure of capture mechanisms that can lead to the generation of more debris objects \cite{Castronuovo2011}. Therefore, it is required that an attitude assessment of defunct objects be performed to ensure suitability for removal. This assessment is best performed before the chaser satellite has spent fuel to intercept the orbit of the target. 

Space debris photometry (light curve analysis) is a widely investigated technique for estimating the attitude dynamics of space debris objects. The light curve of a debris object depends on its attitude, shape, size, and material properties. Inverting light curves to recover these parameters is a notoriously ill-posed problem. Under certain conditions, it is possible to estimate some of these parameters if the others are well known, or if a large quantity of historical measurements are available \cite{Kucharski2021}; however, this typically requires computationally expensive rendering, physics-based orbit and attitude propagation, and can have complicated error surfaces, which makes it difficult to converge on correct solutions \cite{Nussbaum2022}. These solutions become better constrained when data is gathered with higher time-resolution  \cite{Kucharski2019}, incorporates spectral information \cite{Nussbaum2021}, multiple simultaneous observers \cite{Schafer2017}, or when combined with SLR or SDLR data \cite{Nussbaum2021}. However, even when spacecraft are fully characterised, there can still be considerable disagreement between simulated and measured light curves \cite{Meyer2024} which calls into question the accuracy of inversion techniques that rely on this kind of forward-modelling.

A critical limitation of ground-based photometric techniques is that they require the observed object to be sunlit and the observer to be in darkness. This severely reduces observation opportunities. Additionally, obtaining useful information in a single pass requires objects to have fast rotation periods, and the technique becomes difficult if the rotation is slow relative to the duration of a ground station pass, as is often the case with space debris in LEO. In these cases, the state of the object must be propagated between observations with a high-fidelity orbit/attitude simulation \cite{Virdee2017}. Because of these limitations, light curves may not be a suitable technique for providing fully-determined attitude states with low latency, at any time of day, and with modest computational effort, and alternative techniques must be developed.

\begin{figure}
\centering
\includegraphics[width=\linewidth]{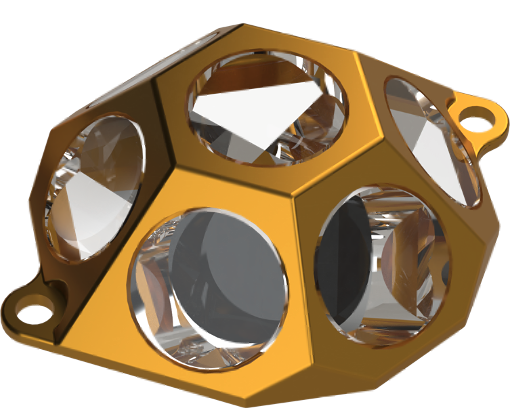}
\caption{The Lumi Micro-2 commercial off-the-shelf retroreflector. These weigh $50$~g and can be placed on the body of a satellite to allow for precision tracking even for defunct satellites. These are designed especially for attitude determination purposes and therefore have an acceptance angle of $160~^o$ for maximum visibility. The placement of three retroreflectors onboard can be used with the methods in this paper to obtain highly precise attitude estimates.}
\label{img: micro 2}
\end{figure}

For cooperative targets, satellite laser ranging (SLR) offers a promising alternative approach for attitude assessment. In this context, a target is considered "cooperative" when it has one or more retroreflectors onboard. Our work focuses exclusively on cooperative targets, specifically those that have been prepared for eventual disposal by including multiple retroreflectors on the satellite body. Fig.~\ref{img: gold retro} shows an example of a commercial off-the-shelf retroreflector, weighing $20$~g, suitable for precise satellite tracking up to $900$~km altitude, the Lumi Micro-1. Laser retroreflectors (LRRs) significantly increase the photons returned to the ground station during SLR. Additionally, they embed attitude information in the photon return signals as they are reflected off a specific known point, moving about the centre of mass, rather than randomly off the satellite body. Fig.~\ref{img: micro 2} shows the Lumi Micro-2 LRR which is designed especially for attitude determination applications and has an acceptance angle of $160~^o$, weighs $50$~g, and can be used up to a maximum altitude of $1300$~km.

Previous SLR-based attitude assessment studies have primarily focused on International Laser Ranging Service (ILRS) geodetic targets, which are typically spherical satellites with numerous retroreflectors arranged at known locations in the satellite's body frame. Researchers have employed frequency analysis on range residuals to infer spin rates for satellites such as Ajisai \cite{Otsubo2000, Kirchner2007, Kucharski2010}, LARES \cite{Kucharski2012}, LAGEOS-1 \cite{Kucharski2007, Kucharski2013}, LAGEOS-2 \cite{Bianco2001, Kucharski2009}, and ETALON-1/2 \cite{Kucharski2008}. These methods rely on spherically-located retroreflector positions relative to the satellite center of mass. Beyond geodetic satellites, attitude assessment has been performed on defunct operational satellites such as ESA's Envisat. However, these methods typically require a flat spin \cite{Wertz1978}. For instance, Pittet et al. \cite{Pittet2017} presented a method for determining Envisat's spin motion using single-pass SLR measurements from a single ground station, but their approach requires multiple rotations to be present in the data to fit the attitude. While highly valuable, these approaches are generally limited to determining rotation of spin-stabilized satellites or those with highly specific geometries.

More recently, Song et al.\cite{Song2024} proposed a novel attitude estimation method using multi-retroreflector differential satellite laser ranging. By simultaneously tracking multiple LRRs on a spacecraft, their method extracts high-precision range difference sequences to determine precise attitude information. However, a limitation, is the inability to uniquely distinguish which LRR each return signal is from. Without this knowledge, there are $12$ valid solutions to a given attitude assessment. Therefore, a method that can distinguish these returns and associate a label to the returns is a necessary next step on the path to unique high-precision attitude solutions.

\begin{figure}
\centering
\includegraphics[width=\linewidth]{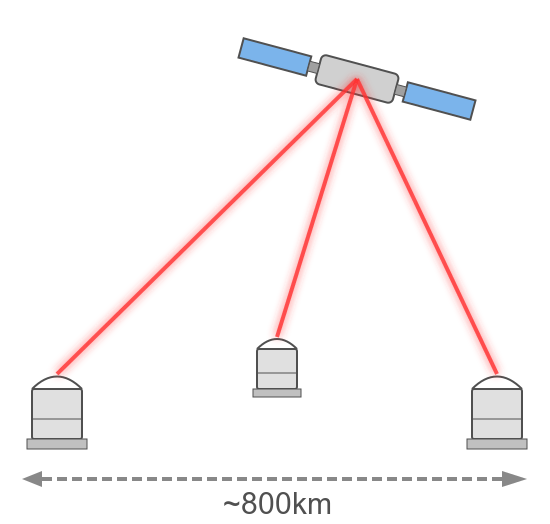}
\caption{Tri-static SLR ground station network concept.}
\label{img: gs layout}
\end{figure}

In this paper, we address these challenges by introducing a tri-static ground-based SLR method for attitude determination, as shown in Fig.~\ref{img: gs layout}. Our approach uses three SLR ground stations making simultaneous observations of a target, each receiving returns from three onboard laser retroreflectors. By capturing range data from different stations and retroreflectors at the same time, we triangulate each retroreflector's position in inertial space. Via this process we can assign labels to the returns from each LRR, therefore matching LRRs in the inertial and body reference frames and allowing for a unique attitude solution to be obtained. This makes our approach operationally robust and provides greater confidence to ADR operators that the attitude estimates they are receiving are correct. Our method does not require any knowledge of the satellite’s mass properties or inertia tensor, relying only on a geometric model of the reflector locations on the spacecraft’s body. We also considered a bi-static approach, which would be less costly and operationally simpler. However, we found it necessary to make assumptions to uniquely label each LRR. Further, even with these assumptions, this approach yielded two possible attitude solutions. By expanding to three stations, we have eliminated these ambiguities, providing only a single attitude solution.

Section~\ref{sec:models and methods} describes our method using multiple stations to provide intersecting range measurements to the satellite’s reflectors, which we use to reconstruct the satellite’s instantaneous orientation with high fidelity.  Section~\ref{sec: extracting angular velocity} shows through simulation that the tri-static approach can determine a satellite’s spin rate and axis with high precision, on the order of one tenth of a degree per second for the spin rate and one degree for the spin axis in the majority of our test cases. Section~\ref{sec:lrr placement} shows results of simulations that provide guidance for satellite manufacturers on optimal placement of multiple LRRs to enable attitude assessment to be performed. Section~\ref{sec: ground station placement} shows results of experiments on ground station placement locations for satellites in LEO. Finally, we conclude in Section~\ref{sec: conclusions}.

\section{Attitude Determination Models and Methods}
\label{sec:models and methods}
Here, we introduce our method for taking simultaneous laser-ranging data from three ground stations to three onboard LRRs and using it to estimate the spin angular velocity of the satellite. We will present a general methodology for triangulating multiple retroreflectors simultaneously and then combining these data together in such a way that the attitude can be extracted. We will also describe the simulation environment and the assumptions that we have made within it.

\subsection{Simulation environment}
To show the performance of this method without an in-orbit demonstration, we have developed a simulation environment such that we can perform attitude determination on simulated data and compare the results to the ground truth data from the simulation itself. In this manner, we can ascertain the likely performance of this approach and show how an in orbit demonstration would likely lead to fruitful results.
We work exclusively with single-shot data from our network and never form normal points with the data as these time average over the rotational motion thereby losing attitude information; while, also reducing the frequency of data points coming into the data pipeline. We use a single-shot precision of $1~$cm throughout this work. Moreover, we assume a return rate of $10$~Hz per LRR per station. For computational efficiency, we assume that each ground station receives a photon simultaneously from each of the LRRs. We have validated this assumption and found the interpolation of ranges between photon returns produces only small errors provided the return rate of the laser is sufficiently high. We do not use a priori information to label the returns from any LRR and this labelling comes naturally through our process. 
 
 We use simulated data through the rest of this article using a Python wrapper around the D-SPOSE \cite{Sagnieres2019} library. We apply an acceptance angle model to each of the LRRs such that we only create simulated ranging returns data when all three LRRs are visible from all ground stations simultaneously. This model is based around the design of the Lumi Micro-2 LRR which gives an almost complete hemisphere of visibility. 
We have removed assumptions from the simulation environment including knowledge of the centre of mass of the satellite, the arrival time of photons, and a priori knowledge of photon returns having come from a specific LRR. We add white Gaussian noise throughout our simulations to account for all of the relevant above effects in addition to the measurement error of the SLR system. 

In reality, there is a two step process for data processing. First, the returns from each LRR must be separated from one another and then these must be passed into the attitude determination pipeline. For computational efficiency, we include work from only with the second step of this process here. However, we do not attach LRR labels to these data points. Instead, the only associated knowledge is that each return is from a different LRR at a given input epoch to the attitude assessment tool. This is a much weaker condition and one that we have satisfied through use of Gaussian mixture models to separate the returns. The centroids in these models can be used for local temporal association of LRRs, e.g. for plane interpolation, but cannot be relied upon for global association as the centroids frequently converge during certain attitude states. 

\subsection{Station and LRR geometry setup}
In an inertial reference frame, we consider three ground stations, denoted $\mathbf{g}_1$, $\mathbf{g}_2$, and $\mathbf{g}_3$, each observing the same three retroreflectors located on the body of the satellite. The unknown ground-truth retroreflector locations in the inertial frame are denoted as $\mathbf{l}_1^{i}$, $\mathbf{l}_2^{i}$, and $\mathbf{l}_3^{i}$. In contrast, in the body frame of the satellite the retroreflector locations are known and in this coordinate system we denote these as $\mathbf{l}_1^{b}$, $\mathbf{l}_2^{b}$, and $\mathbf{l}_3^{b}$. In both cases, the superscript refers to the inertial or body reference frame. Henceforth, we refer to the locations of the LRRs and their normal vectors in the body frame as the satellite geometric model. Our technique works by first using triangulation methods to ascertain possible locations of the LRRs in inertial space and then finding the permutation of LRR placement within these possible locations that best matches the satellite geometric model.

To find potential locations, we take range measurements from each of the three stations to each of the LRRs simultaneously. In practice, the return time of the photons and therefore the measurement epoch will be slightly different for each station. We have performed experiments to ascertain the interpolation error expected to interpolate the ranges to the same epochs to perform our method. We find the error is quite small but include it as noise in our range measurements for completeness. 

Each ground station will obtain a time series of range measurements and a priori there is no way to know which LRR a range measurement has come from despite this being essential to performing attitude determination on these data. These range time series are denoted as $\mathbf{r}_1$, $\mathbf{r}_2$, and $\mathbf{r}_3$ where the index refers to the ground station number and each vector has a length equal to the number of range measurements taken by a particular ground station. Note that it is assumed that the SLR station is operating at the single-photon level and that no first photon bias effects are present across the three LRRs. 

We therefore have three ranges from each ground station that correspond to the three onboard LRRs. This technically corresponds to three spheres centered at the ground station with a radius equal to the three ranges themselves. However, we also have an approximate pointing direction of the lasers during the range measurement from the azimuth and elevation coordinates taken from the tracking mount. Using this information, it is sufficient to approximate the local portion of each sphere by a plane passing near the satellite. 

If, for now, we assume that we can separate the returns from the three LRRs from a single ground station, then we have three planes, one approximately passing through each LRR with a normal vector pointing along the laser. We have these three planes for each ground station, therefore totalling nine planes. The intersection of these nine planes with one another leads to a total of twenty-seven vertices where it is possible that an LRR is located. We refer to these as $$V = \{\mathbf{v}_1,\mathbf{v}_2,...,\mathbf{v}_{27}\}$$ where $\mathbf{v}_i$ is a vector corresponding to the location in inertial space of the $i_\textrm{th}$ vertex, and we call these LRR candidate locations. It should be noted that the location of these vertices has an error associated with it due to the single-shot precision of the SLR measurement and also due to the geometry of the ground stations relative to each other and to the satellite itself (see Sec.~\ref{sec: ground station placement} for a metric on the expected triangulation precision based on these factors). Therefore, each of the vertices should not be viewed as a true LRR candidate location but as a statistical estimate of the location sampled from a distribution with a covariance depending on the aforementioned factors. When using the previously discussed laser configuration these samples typically have a triangulation error of approximately $3$~cm. A key hurdle to overcome with the development of this algorithm was ensuring that we are able to accurately determine the attitude in the presence of these errors. 

\subsection{Locating the LRRs in inertial space}
Through the process of triangulating three LRRs simultaneously from three ground stations and then finding the intersection of the resulting nine planes we can infer twenty-seven LRR candidate locations in inertial space. There are three LRRs in any permutation of twenty-seven locations meaning a total of $27\times26\times25 = 17550$ possible permutations $P$ to assess to find which permutation best fits against the known satellite geometric model, i.e. the locations and normal vectors of the LRRs in the body frame. Note that including four LRRs on the satellite body results in $64\times63\times62\times61 > 15\times10^6$ permutations to assess which, as will be seen, is highly computationally expensive. It is partially for this reason that we recommend three and not more LRRs visible at any one time for attitude assessment.

We use two loss functions to assess the LRR candidate locations against the geometric model. The first loss function, $\mathcal{L}_1$ (Sec.~\ref{sec: L1}), is much faster to calculate, and we therefore run it on all $17,550$ possibilities. The second loss function, $\mathcal{L}_2$ (Sec.~\ref{sec: L2}),  is slower but gives better estimates so we only apply it to the top $100$ candidates as defined by the first loss function. We find the best performance if one only accepts a candidate permutation if both loss functions return the same candidate. This process does lead to more true positive candidate selections being thrown away than otherwise but as shown in Sec.~\ref{sec: Filtering of quaternion time series estimates} this is a worthwhile trade off for reducing false positives. Application of these loss functions yields the location of each specific LRR in inertial space and matches the retroreflectors in inertial space against the LRRs in the body frame model allow for attitude assesment. As far as we are aware, this has not been done systematically in the literature previously and is a key contribution of this work.

\subsubsection{Definition of the First Loss Function \(\mathcal{L}_1\)}
\label{sec: L1}
Using the known retroreflector locations in the body frame \(\mathbf{l}_1^b, \mathbf{l}_2^b, \mathbf{l}_3^b\), define the constant separation vector for a given satellite
$$
\delta L^b
=
\begin{bmatrix}
\|\mathbf{l}_1^b - \mathbf{l}_2^b\| \\[6pt]
\|\mathbf{l}_2^b - \mathbf{l}_3^b\| \\[6pt]
\|\mathbf{l}_3^b - \mathbf{l}_1^b\|
\end{bmatrix}.
$$

From triangulation, there are 27 candidate locations \(V = \{\mathbf{v}_1,\dots,\mathbf{v}_{27}\}\) in inertial space. A permutation \(p \in P\) selects one triplet \(\bigl(\mathbf{v}_{p(1)},\mathbf{v}_{p(2)},\mathbf{v}_{p(3)}\bigr)\). For each \(p\), define the inertial‐frame separation vector
$$
\delta L^i_p
=
\begin{bmatrix}
\|\mathbf{v}_{p(1)} - \mathbf{v}_{p(2)}\| \\[6pt]
\|\mathbf{v}_{p(2)} - \mathbf{v}_{p(3)}\| \\[6pt]
\|\mathbf{v}_{p(3)} - \mathbf{v}_{p(1)}\|
\end{bmatrix}.
$$

We then define the total error for permutation \(p\) by taking the norm of the difference between these two vectors
$$
\mathcal{L}_1(p)
=
\bigl\|\delta L^i_p - \delta L^b \bigr\|.
$$

After computing \(\mathcal{L}_1(p)\) for all permutations \(p \in P\), we choose the permutation
$$
p^*
=
\underset{p \in P}{\arg\min}
\;
\mathcal{L}_1(p)
$$
as a highly likely mapping from the candidate locations to the geometric model. We have found this method alone to be remarkably reliable on simulated data provided that certain geometric conditions on retroreflector placement, discussed later, are observed.

\subsubsection{Definition of the Second Loss Function \(\mathcal{L}_2\)}
\label{sec: L2}
As before, let \(p \in P\) index the candidate triplet 
\(\bigl(\mathbf{v}_{p(1)},\mathbf{v}_{p(2)},\mathbf{v}_{p(3)}\bigr)\)
in inertial space. Denote the three known vertices of the retroreflector triangle in the body frame by
\(\mathbf{l}_1^b, \mathbf{l}_2^b, \mathbf{l}_3^b\), and collect them in a matrix
\(\mathbf{T}\in\mathbb{R}^{3\times 3}\). The corresponding candidate triangle is
\(\mathbf{C}\in\mathbb{R}^{3\times 3}\), formed by the rows
\(\mathbf{v}_{p(1)},\mathbf{v}_{p(2)},\mathbf{v}_{p(3)}\).

Kabsch's method is applied to find the optimal rotation and translation of \(\mathbf{P}\) onto \(\mathbf{T}\) after which a sum of squared distances error metric is applied. If \(\mathbf{R}\) is the optimal rotation matrix then the loss is given by

$$
\mathcal{L}_2(p)
\;=\;
\sum_{i=1}^3
\Bigl\|
\bigl(\mathbf{l}_i^b - \bar{\mathbf{l}}^b\bigr)
\;-\;
\bigl(
\mathbf{R}\,\bigl(\mathbf{v}_{p(i)} - \bar{\mathbf{v}}\bigr)
\bigl)
\Bigr\|^2
$$

where \(\bar{\mathbf{l}}^b\) and \(\bar{\mathbf{v}}\) are the respective centroids of \(\{\mathbf{l}_i^b\}\) and \(\{\mathbf{v}_{p(i)}\}\). In other words, \(\mathbf{R}\) is chosen to minimize the sum of squared point‐to‐point distances between the shifted triangles. Once \(\mathcal{L}_2(p)\) is computed for all permutations \(p \in P\), a highly likely placement of the retroreflectors is determined by

$$
p^*
\;=\;
\underset{p\in P}{\arg\min}\;\mathcal{L}_2(p).
$$

We find that this method more reliably predicts the correct LRR locations than $\mathcal{L}_1$, often by approximately $5-10\%$, but also that it is more computationally expensive. 

\begin{figure}[]
\centering
\begin{subfigure}{\columnwidth}
  \centering
  \includegraphics[width=\linewidth]{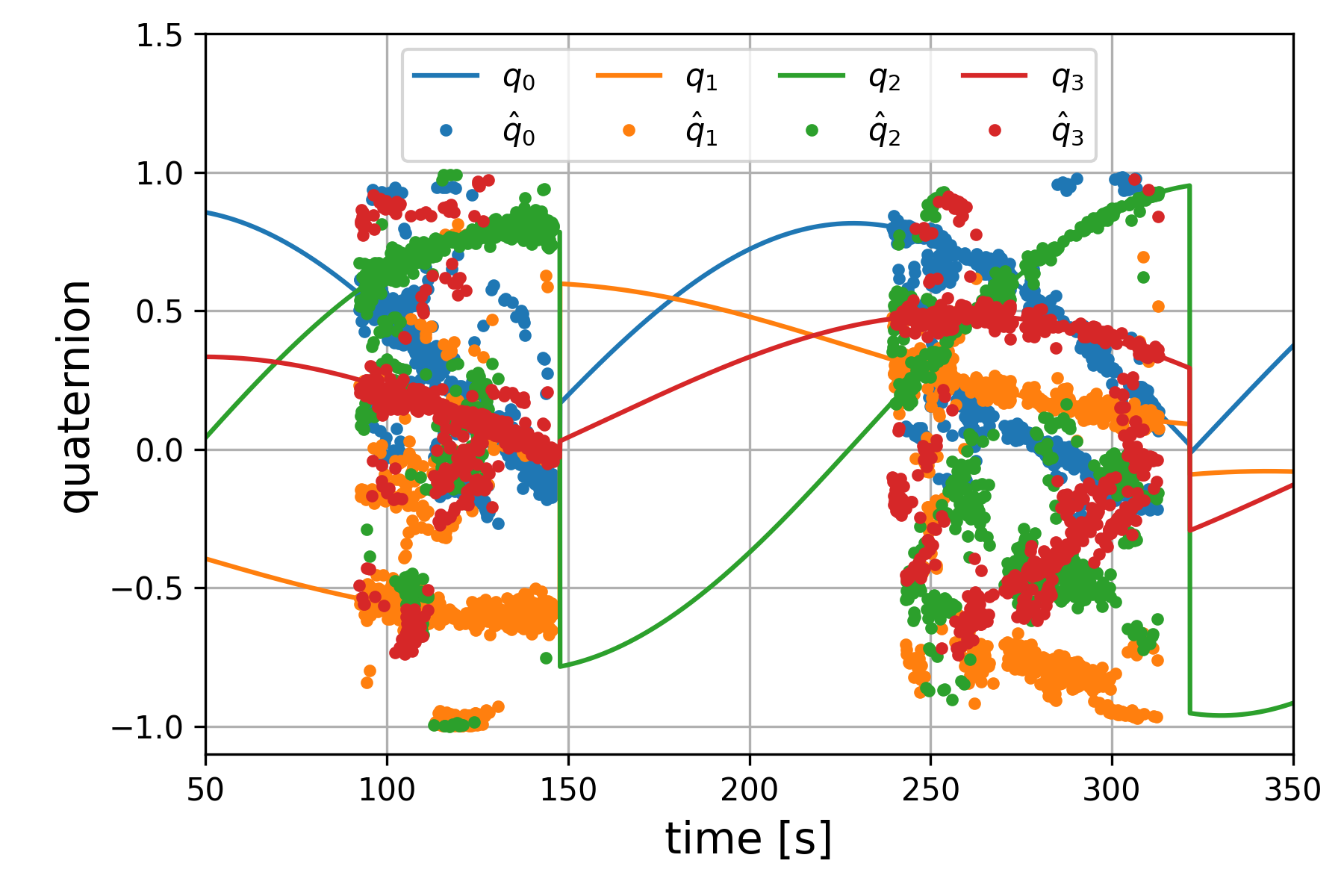}
  \caption{Output of quaternion time series estimation process at all points in time without data rejection.}
  \label{fig: q after l2 loss}
\end{subfigure}

\bigskip

\begin{subfigure}{\columnwidth}
  \centering
  \includegraphics[width=\linewidth]{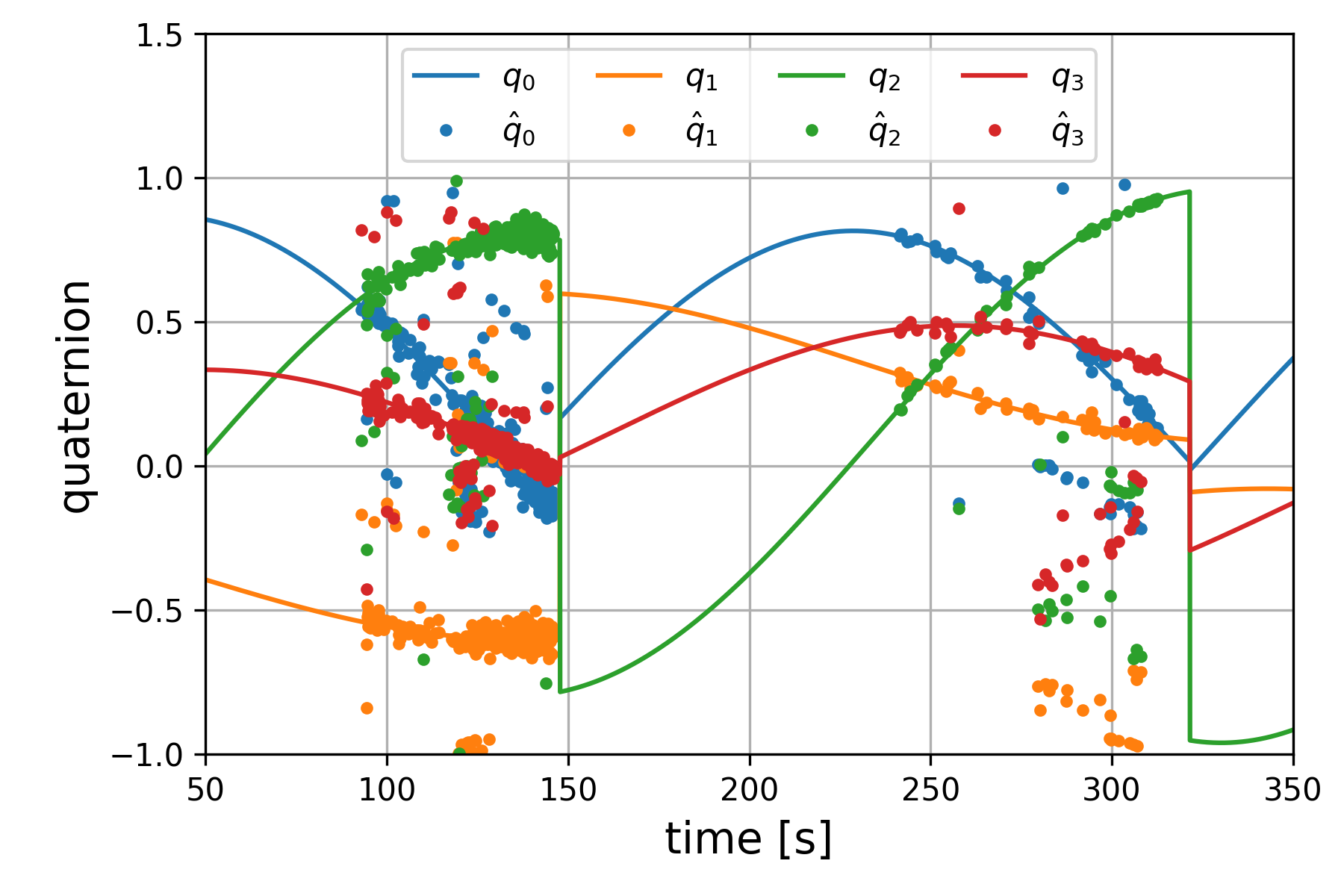}
  \caption{Output of quaternion time series estimation process at all points in time with loss-function-based data rejection.}
  \label{fig: q after loss functions}
\end{subfigure}

\bigskip

\begin{subfigure}{\columnwidth}
  \centering
  \includegraphics[width=\linewidth]{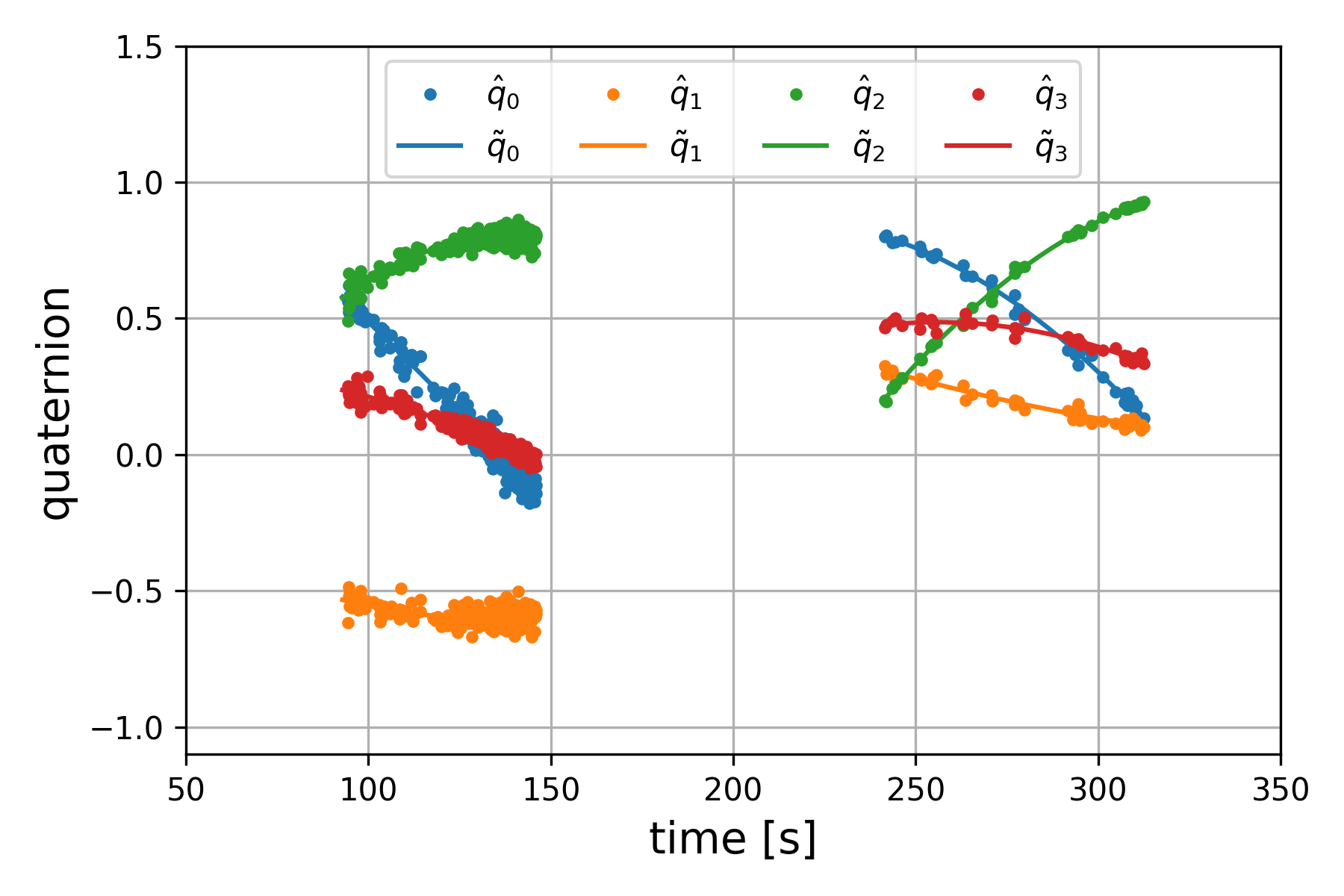}
  \caption{Output of quaternion time series estimation process after polynomial fitting and final outlier rejection.}
  \label{fig: q after polynomial fitting}
\end{subfigure}

\caption{Time series estimates of the satellite attitude quaternion at various stages of the attitude determination method data processing pipeline. The ground truth quaternion being estimated is shown as solid lines for the four components. The corresponding quaternion estimates are shown as dots for each data point estimated via multi-static triangulation.}
\label{fig:quaternion_estimation}
\end{figure}

\subsection{From LRR locations to an attitude quaternion time series}
Given a selected permutation, $p^*$, we can now use these data to obtain an attitude quaternion. First, we calculate the centre of mass of the satellite in the inertial frame so that we have an origin in both frames to perform our rotations about. To do this, we once again use Kabsch's algorithm to find the best rotation and translation that minimises the error between the geometric model LRRs and the LRRs in inertial space as contained by $p^*$. The centre of mass in the inertial frame is then fully defined by the geometric model with the reflection case being handled by the LRR acceptance angle model. Note that this is an estimate of the centre of mass and contains errors. Using simulated data we find that these errors are small and much less than the precision of the laser in most cases. However, for completeness we propagate these errors throughout the simulation by adding them to the estimated LRR locations before calculating the attitude quaternion. We then solve the over-defined system of equations containing the LRR locations in inertial and body frames to find the attitude rotation matrix estimate. This matrix is used to obtain an estimate of the attitude quaternion, $\hat{\mathbf{q}}$.  We then repeat this process at all measurement epochs to obtain a time series of quaternion estimates.

\subsection{Filtering of quaternion time series estimates}
\label{sec: Filtering of quaternion time series estimates}

\begin{figure}
\centering
\includegraphics[width=\linewidth]{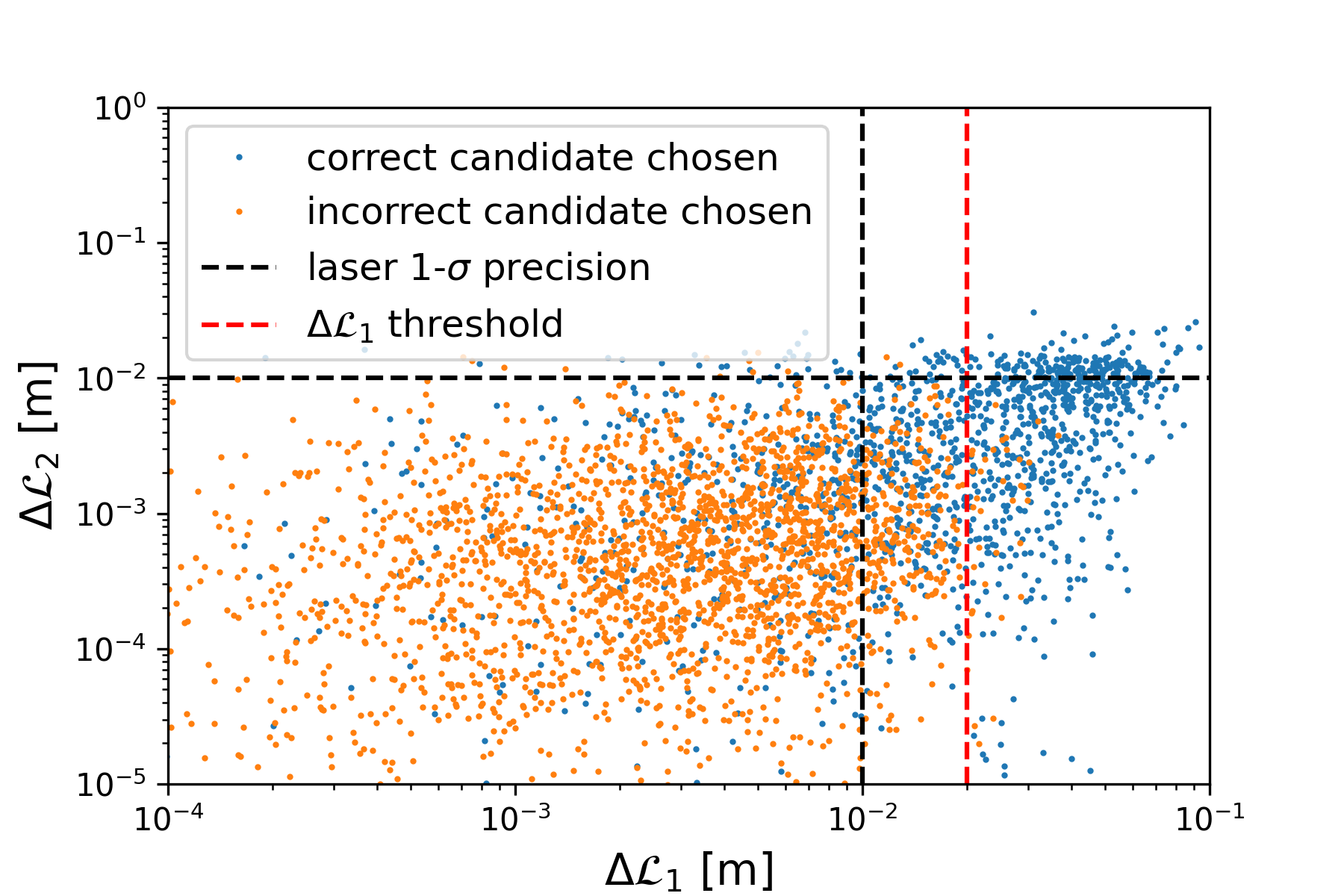}
\caption{The difference between the primary and secondary LRR placement permutations effect on correct selection. The x-axis shows this for $\mathcal{L}_1$. The y-axis shows this for $\mathcal{L}_2$. The blue dots indicate the times that the $\mathcal{L}_1$ loss function identified the permutation correctly whereas the orange data points indicate an incorrect selection. The $1$-$\sigma$ precision of the laser is shown as a dashed black line and the $2$-$\sigma$ threshold used for data rejection is shown as a dashed red line.}
\label{img:l1_l2_diff}
\end{figure}

Fig.~\ref{fig:quaternion_estimation} contains the results of a quaternion estimation process for a satellite pass over a ground station, for this experiment we choose favourable but realistic conditions on pass geometry and LRR placement in accordance with the findings in Sec.~\ref{sec:lrr placement}~and~\ref{sec: ground station placement}. The spacecraft is rotating at $2~^os^{-1}$ and the gaps in data are caused by the attitude of the rotating space craft meaning the line of sight from one or more ground stations to the onboard LRRs is blocked. The ground-truth quaternion, known only due to this being a simulation, is shown as four solid lines whereas the estimated quaternions, $\hat{\mathbf{q}}$, are shown as dots. Fig.~\ref{fig: q after l2 loss} is the quaternion estimate if we simply trust the output of the $\mathcal{L}_2$ loss function. We see that although a lot of data points are correct we also find a lot of outliers. In this particular case, we find that the $\mathcal{L}_1$ loss selects the correct candidate $31.7\%$ of the time and the $\mathcal{L}_2$ loss selects correctly $36.3\%$ of the time. If we consider the case that either one of them select the correct candidate then this percentage increases to $46.4\%$; clearly, there is a benefit to the combination of both metrics. Finally, in the case that both metrics must select the same candidate in order for the candidate to be accepted the total data points count is reduced to $26.3\%$ of the original sample size. However, of these remaining candidate selection $82.2\%$ of them are true positives. We would like to maintain at least one sample per degree of rotation of the satellite, but the sample data frequency  depends upon the laser parameters and is not fully considered here. Nor have we considered a wide parameter space of satellite rotations. However, retention of only $26\%$ of the overall sample will be ample for these goals based upon expected laser parameters and spin rates typical of ADR planning. Therefore, we simply state that can afford to lose samples in our pipeline if it means we can simultaneously reduce the number of false positives, and for this reason we use this criterion of both metrics identifying the correct candidate despite the loss of $73.6\%$ of the total sample size. Despite this, we will now apply another method to retain some more discarded points.

Fig.~\ref{img:l1_l2_diff} shows the difference between the candidate with the lowest loss, $p^*$, and the candidate with the second lowest loss. This difference is shown for the two loss functions, $\mathcal{L}_1$ and $\mathcal{L}_2$, and we denote them as $\Delta \mathcal{L}_1$ and $\Delta \mathcal{L}_2$, respectively. This difference makes sense as a useful metric as differences between primary and secondary candidates that are less than the precision of the laser are likely indistinguishable from one another. In Fig.~\ref{img:l1_l2_diff}, where the blue dots indicate the correct candidate being chosen by $\mathcal{L}_1$ and the orange indicate a false candidate selection. The laser precision is shown in black and the plot shows that the $\Delta \mathcal{L}_1$ metric delineates well the true and false positives about the laser precision value. We see no such delineation when using the $\Delta \mathcal{L}_2$ metric. As such, we have included a threshold at the laser $2\sigma$ threshold such that any data points that have a $\Delta \mathcal{L}_1 > 2\sigma$, where $\sigma$ is the single-shot laser precision, are also accepted. This threshold is shown as a dashed red line on the plot. This corresponds to a true positive candidate selection of $98.8\%$ with a loss of $53.7\%$ of the available data points. Therefore, the final metric we use for data inclusion is that both loss functions return the same prediction or that $\Delta \mathcal{L}_1 > 2\sigma$.

Fig.~\ref{fig: q after loss functions} shows the quaternion estimation data points, $\hat{\textbf{q}}$, remaining after the two loss functions have been applied and the additional $\Delta \mathcal{L}_1$ metric has been used for filtering. Here, the remaining data points are much more tightly clustered about the ground truth quaternion elements shown in solid lines. However, there are still some outliers present and the final step in the data reduction process is to fit a piecewise polynomial to each section of data points and to apply iterative outlier rejection. We do this with a least squares algorithm to fit a second-order polynomial using the trust region reflective algorithm with a Huber loss function. We iterate this process pruning outliers greater than a standard deviation until we are no longer removing points. 

Fig.~\ref{fig: q after polynomial fitting} shows the remaining data points after this final outlier rejection process where we can see a much tighter distribution of remaining data points. The interpolated quaternion, $\tilde{\boldsymbol{q}}$, is shown and it is this interpolated quaternion that is now used to determine the spin angular velocity time series.

\subsection{Extracting angular velocity}
\label{sec: extracting angular velocity}
\begin{figure}
\centering
\begin{subfigure}{\columnwidth}
  \centering
  \includegraphics[width=\linewidth]{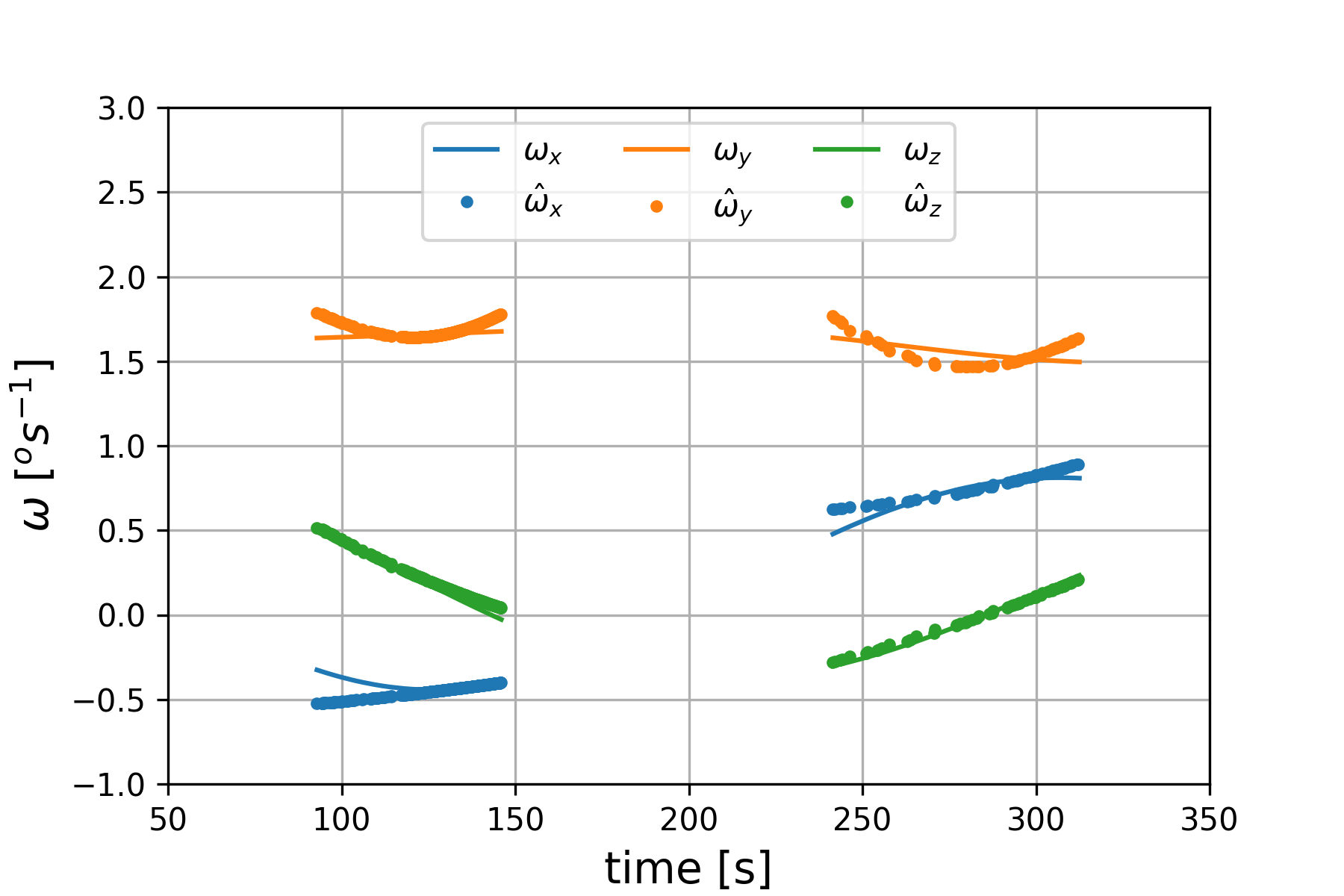}
  \caption{Spin angular velocity component estimates. The ground truth angular velocity from the attitude determination process is shown component wise as three solid lines. The estimated angular velocity is shown as dots. Colour represent directional components.}
  \label{fig: angular velocity estimate}
\end{subfigure}

\bigskip

\begin{subfigure}{\columnwidth}
  \centering
  \includegraphics[width=\linewidth]{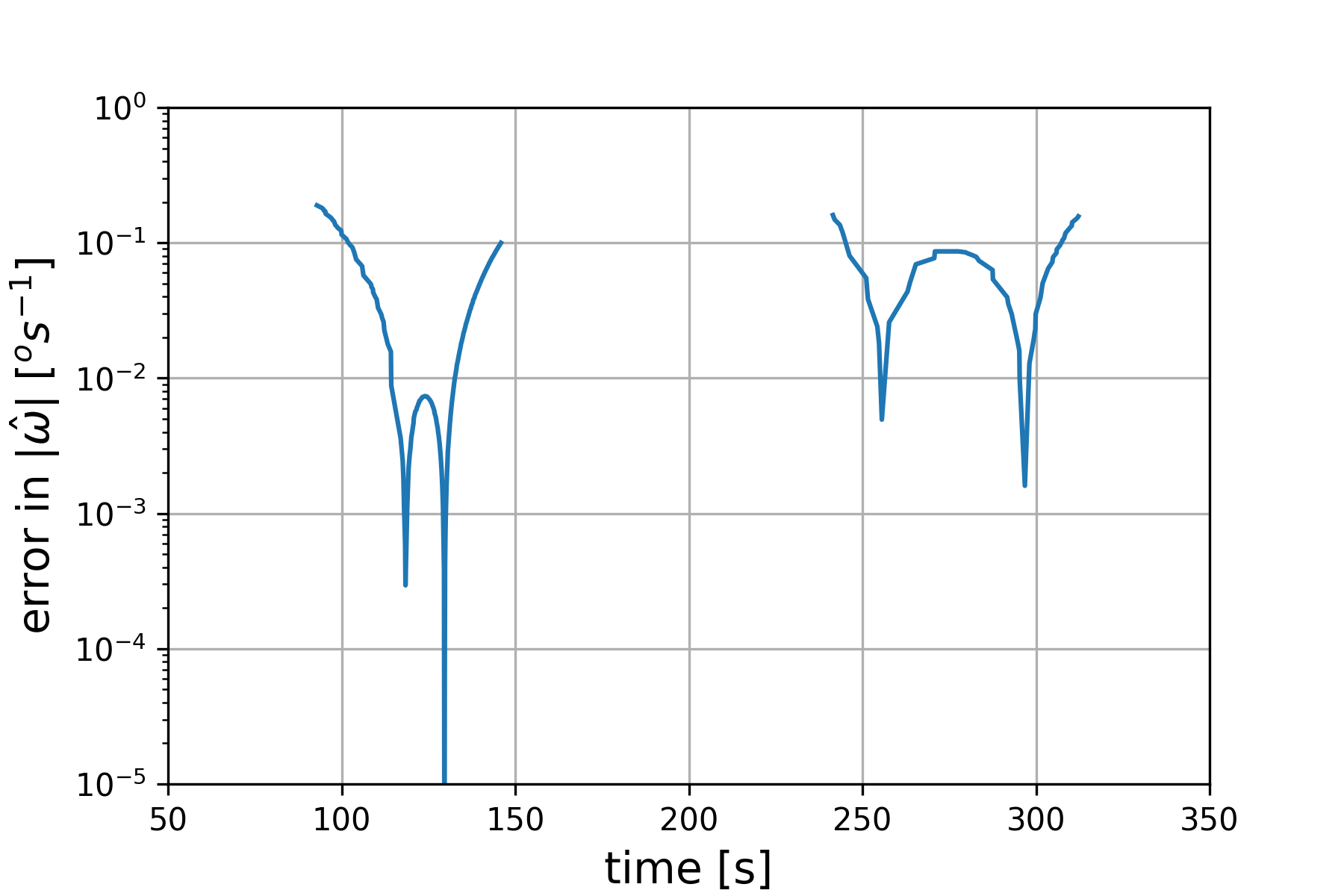}
  \caption{The error in the estimated angular velocity shown in the panel above when only the magnitude of the overall vector is considered.}
  \label{fig: angular velocity error}
\end{subfigure}
\caption{Time series estimates and errors of the satellite spin angular velocity obtained via the tri-static SLR-based attitude determination method.}
\label{fig:angular velocity estimation}
\end{figure}

Given our polynomial interpolation, $\tilde{\boldsymbol{q}}$, of the quaternion estimates, $\hat{\mathbf{q}}$, we can now extract the angular velocity using a finite difference scheme of adjacent quaternions. We find this approach to be a more stable then differentiating the polynomials themselves directly. The polynomials are used to obtain a time series of quaternions at $1$ second intervals and successive quaternions are used in a finite differencing scheme to obtain a time series of angular velocity vector estimates. Given two successive quaternions, $\mathbf{q}_1$ and $\mathbf{q}_2$ that represent attitudes at times $t_1$ and $t_2 = t_1 + \Delta t$, the angular velocity estimate in the body frame, $\hat{\boldsymbol{\omega}}$, is given by 
$$
\hat{\boldsymbol{\omega}} = \dfrac{2}{\Delta t} (\textrm{Im}(\mathbf{q}_1^{-1} \otimes \mathbf{q}_2)),
$$
where $\textrm{Im}(\cdot)$ denotes the vector part of the quaternion. This is applied to all times in our time series to obtain the angular velocity vector at all times.

Fig.~\ref{fig: angular velocity estimate} shows the final output of this process where each of the components, $\hat{\omega}_x$, $\hat{\omega}_y$ and $\hat{\omega}_z$, of the angular velocity estimate, as a result of the processing through the entire pipeline, are presented alongside the ground-truth angular velocity vector components, $\omega_x$, $\omega_y$, and $\omega_z$. The fit between estimated and ground-truth angular velocities is remarkably close and shows the capability of attitude determination via multi-static SLR measurements. The absolute precision of this method is highlighted in Fig.~\ref{fig: angular velocity error} where the error in the angular velocity magnitude is plotted for this example and is shown to be less than a tenth of a degree per second across the vast majority of the attitude estimation process and reaches an error level over an order of magnitude lower during parts of the process. Given that for active debris removal mission planning one needs only acquire a single value of the angular velocity over a given observation it is also possible to use a median across the time series to provide a better value than any individual data point.

\section{Laser Retroreflector Placement}
\label{sec:lrr placement}

\begin{figure*}[ht]
\centering
\begin{tabular}{cc}
  \includegraphics[width=\columnwidth]{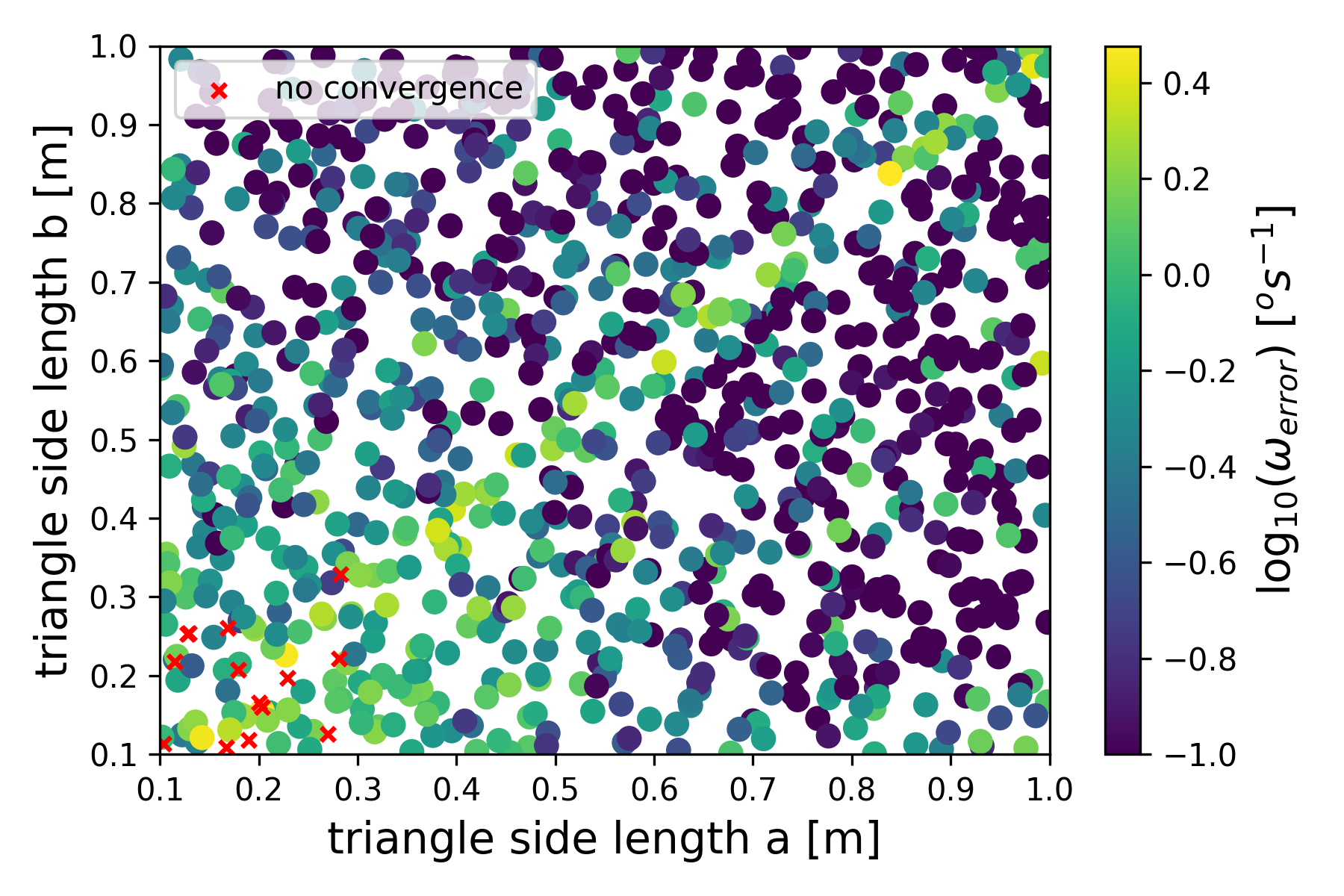} &
  \includegraphics[width=\columnwidth]{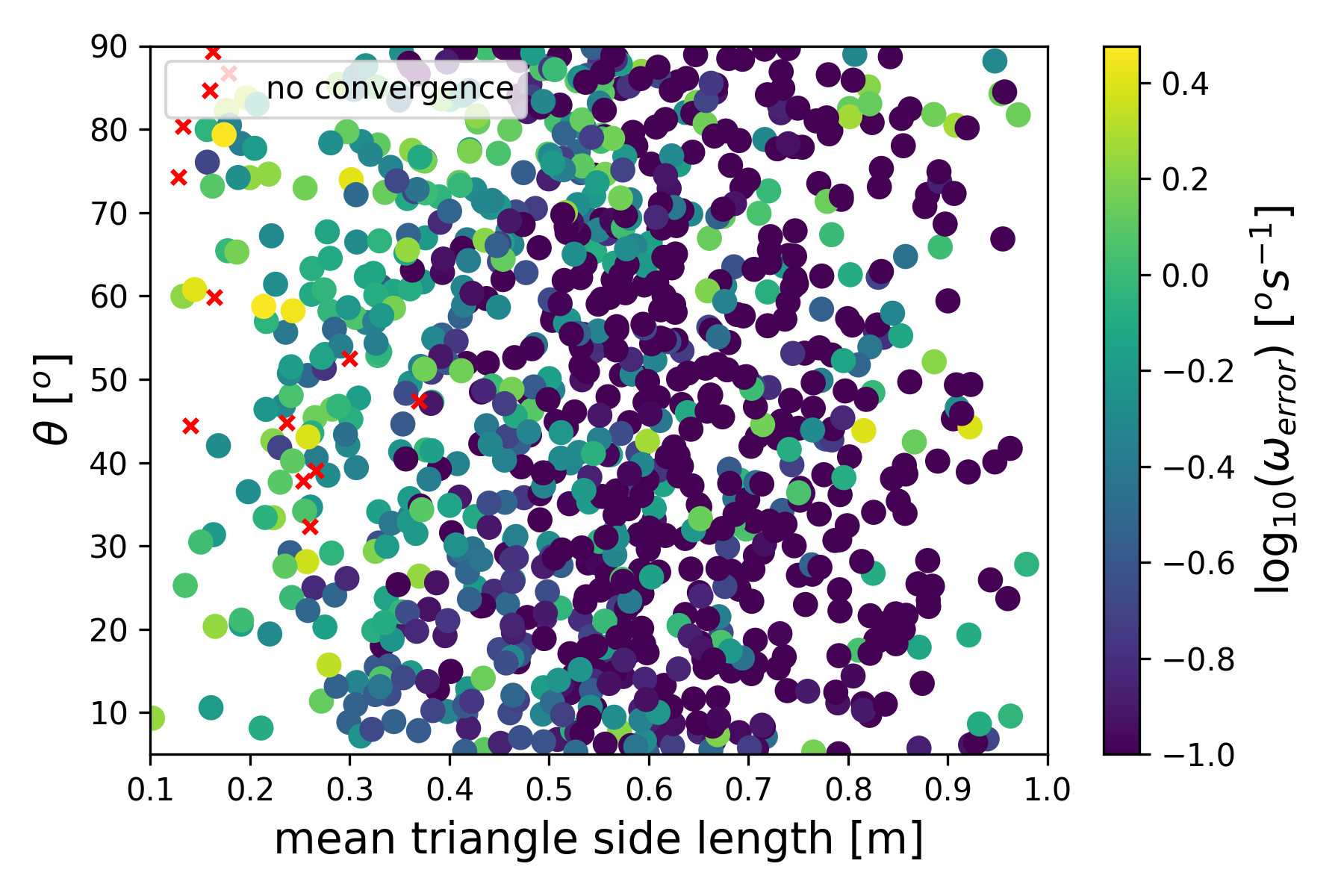} \\
\end{tabular}
\caption{Comparison of retroreflector placement location choice considering a triangle layout geometry via the two side lengths, $a$ and $b$, and angle $\theta$. Each data point represents the output of an attitude determination simulation realisation. The colour indicates the log of the median error in the angular velocity magnitude estimation time series for a given run. Red crosses indicate that the AD system did not converge. These plots are key for operators to be able to correctly choose locations on their satellites for placing retroreflectors.}
\label{fig:lrr placement locations}
\end{figure*}

\begin{figure*}[ht]
\centering
\begin{tabular}{cc}
  \includegraphics[width=\columnwidth]{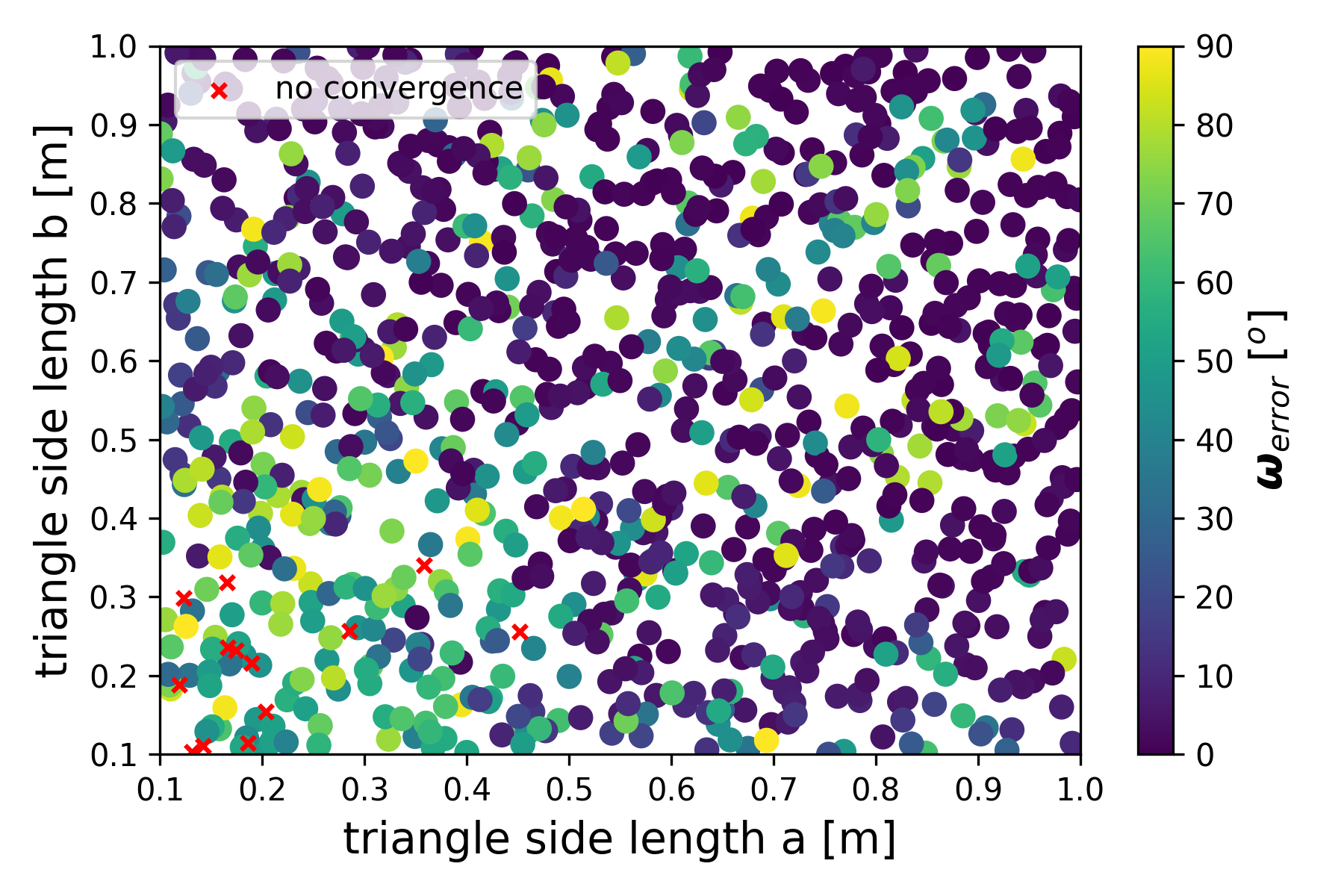} &
  \includegraphics[width=\columnwidth]{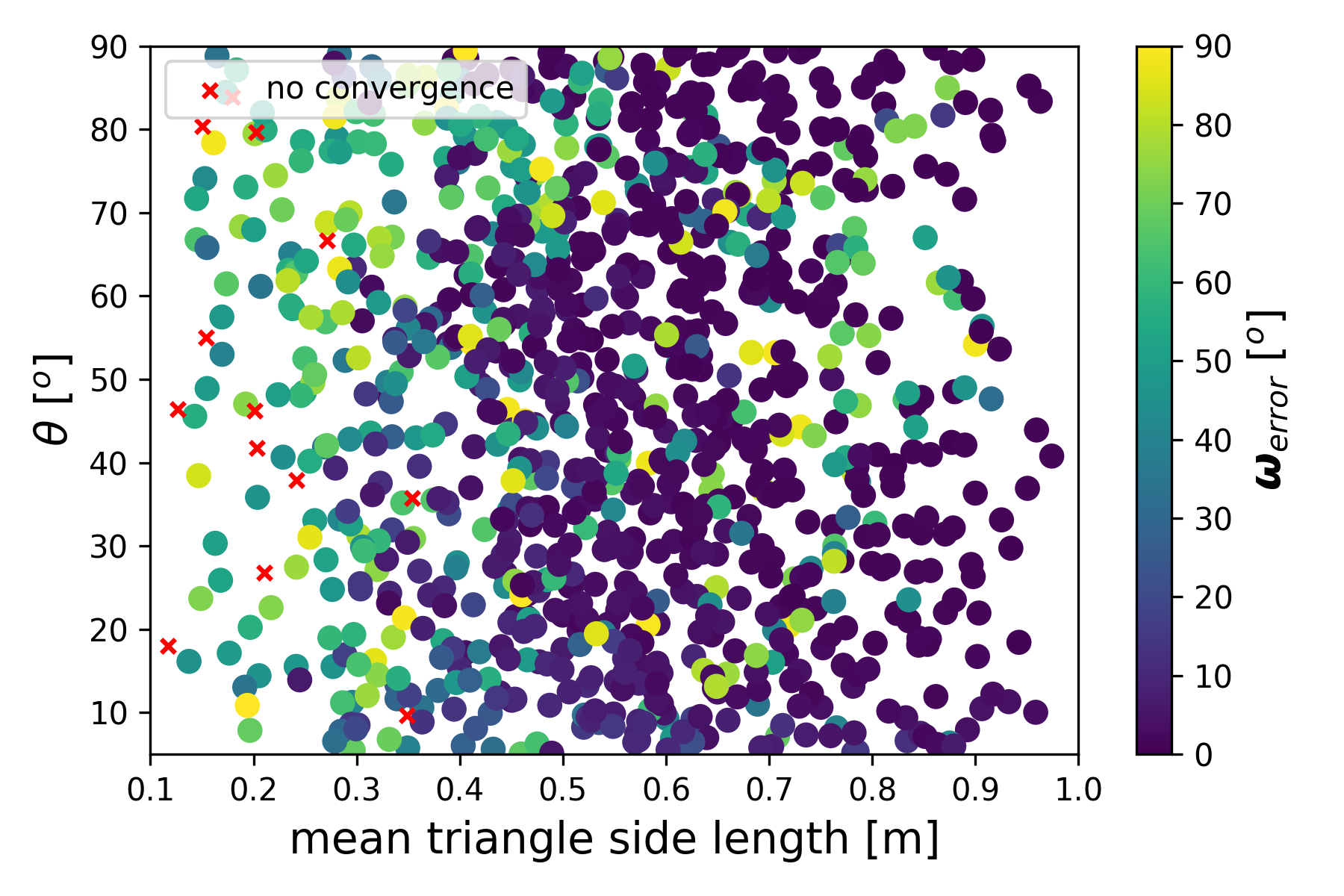} \\
\end{tabular}
\caption{Comparison of retroreflector placement location choice considering a triangle layout geometry via the two side lengths, $a$ and $b$, and angle $\theta$. Each data point represents the output of an attitude determination simulation realisation. The colour indicates the median pointing error in the angular velocity vector estimation time series for a given run. Red crosses indicate that the AD system did not converge. These plots are key for operators to be able to correctly choose locations on their satellites for placing retroreflectors.}
\label{fig:lrr placement locations vector}
\end{figure*}
The goal of this work is to lay the foundations for spacecraft operators to behave more sustainably through enabling end-of-life operations. One key piece of information for operators to enable easy preparation of their satellites for laser-based attitude determination is in what layout should they place their three retroreflectors to maximise performance.
We have conducted a large-scale simulation campaign to investigate the technique’s sensitivity to retroreflector placement, and to develop clear guidelines for satellite operators who wish to include this capability on their spacecraft. For particular mission requirements and placement constraints, we welcome interested parties to contact the authors and Lumi Space, for support in choosing the correct placement for their satellites to maximise attitude determination performance.

The placement of LRRs on a satellite body is a problem with a large parameter space of configurations. First, we shrink this search space by stating that three LRRs are best placed on a single side of the satellite and mounted such that their normal vectors are parallel, i.e. mounted on the same flat surface. LRRs should be placed to minimise obscuring lines of sight where possible. The problem then becomes a case of choosing the optimal geometric layout of the LRRs which can be posed as finding the optimal triangle with an LRR at each vertex. Therefore, the problem can be parametrised by three variables, $a$, $b$, and $\theta$ which are two triangle side lengths and the angle of the vertex between them. Ergo, we can perform a search over these three variables and perform an attitude determination simulation to see the errors obtained by the process in the aforementioned manner. 

In this section, we only consider a single pass at $1200$~km altitude over three ground stations spaced as an equilateral triangle with vertices placed approximately $1000$~km from the triangle centroid. We consider a satellite spin rate of $2.5$~$^o$s$^{-1}$ rotating around all principal axes. We perform $1000$ attitude determination runs for the same pass each and vary only $a$, $b$, and $\theta$ using latin hypercube sampling over the parameter space to maximise coverage. We collect all performance metrics from our attitude determination runs and generate summary statistics to ascertain which configurations of retroreflector placement yield the best performance.

Fig.~\ref{fig:lrr placement locations} contains a scatter plot of the results of this experiment in terms of the median angular velocity error across all points in the time series, $\omega_{error}$. In both plots, this is shown as the colour of the data point and is log scaled over a range of $0.1$ to $3~^o\textrm{s}^{-1}$. The dozen or so runs that did not converge due to the LRR placement triangle dimensions being comparable to the laser precision are marked with a red cross. 

In the left hand panel, this error is plotted against the two parameterised triangle side lengths and clear patterns can be seen. Firstly, there is a distinct region of poor performance where both triangle side lengths are small that gradually improves as the lengths increase. However, it should also be noted that there is a region of poor performance across the diagonal of the heatmap where both triangle side lengths are identical. This is caused by the symmetric shape of the triangle leading to an inability for the geometic model to be used for labelling each retroreflector correctly. Therefore, we suggest that operators avoid equilateral and isosceles triangles in their retroreflector placements and use the heatmaps to ensure that the difference between triangle side lengths is suitably large. Finally, we can see regions of excellent performance on these plots reaching precision below $0.1~^o\textrm{s}^{-1}$ and these are characterised by large triangles, on the order of $>0.8$~m for $a$ and $>0.3$~m for $b$ provided equilateral and isosceles layouts are avoided. 

In the right panel, the same error is shown but plotted against the angle between the sides of the triangle, $\theta$, and the mean triangle side parameter length, $\dfrac{1}{2}(a+b)$. This plot shows less of a dependency on the particular value of $\theta$ on the performance although there is a slight preference for lower values of $\theta$ if the average triangle side length is small. This means that there are regions of good performance across lots of different triangle configurations thereby offering a lot of flexibility to satellite designers and operators when trying to accommodate three LRRs on their satellite.

Fig.~\ref{fig:lrr placement locations vector} contains results from the same experiments runs as Fig.~\ref{fig:lrr placement locations}. Here, the output variable being plotted is the median time series angular error in the angular velocity vector, named $\boldsymbol{\omega}_{error}$. The error is not log-scaled as in the previous plots but instead is linear over the range of $0$ to $90$ degrees. We find a similar pattern to the previous results on the angular velocity magnitude error and unsurprisingly find that runs that perform well in this regard also perform well in terms of angular error. With good retroreflector placement, we find that an error of approximately $1~^o$ can be obtained in the majority of simulated runs. We observe that even in the regions of excellent performance there are runs that perform worse than the others nearby in the input parameter space, these are rare and are a result of the specific noise conditions of the run. We have made no concerted effort to improve the numerics to enable a better convergence on these runs but looking at the data it seems that these errors can be reduced in the future, in particular once the return rate of the laser is better quantified. This has been left for future work along with a means of identifying these outliers.



\section{Ground Station Placement}
\label{sec: ground station placement}
\begin{figure*}
\centering
\begin{tabular}{c}
  \subcaptionbox{Inclination $53^o$
   \label{fig:total passes (a)}}{%
    \includegraphics[width=\textwidth]{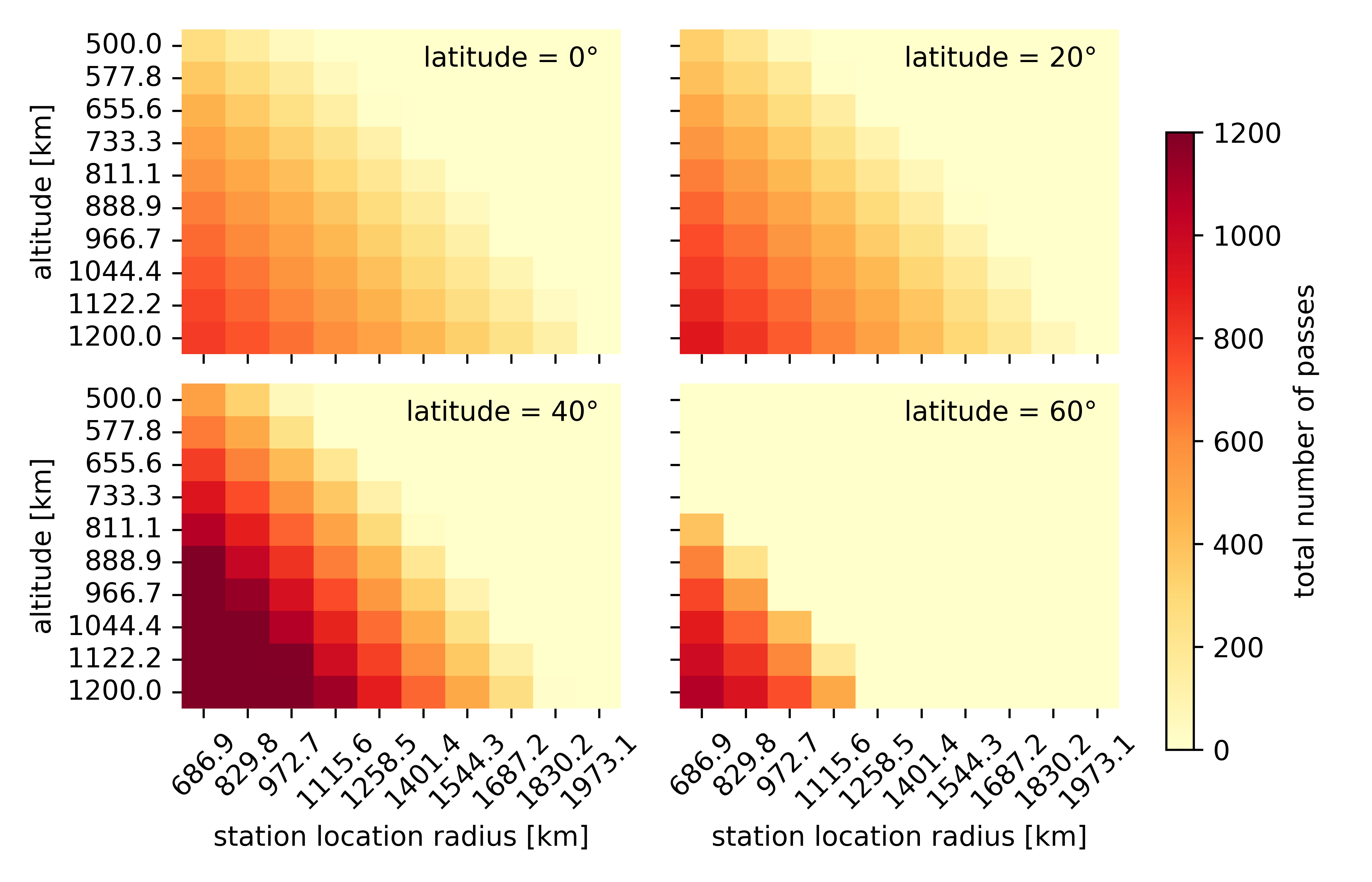}
  }\\[1ex]
  \subcaptionbox{Inclination $86^o$\label{fig:total passes (b)}}{%
    \includegraphics[width=\textwidth]{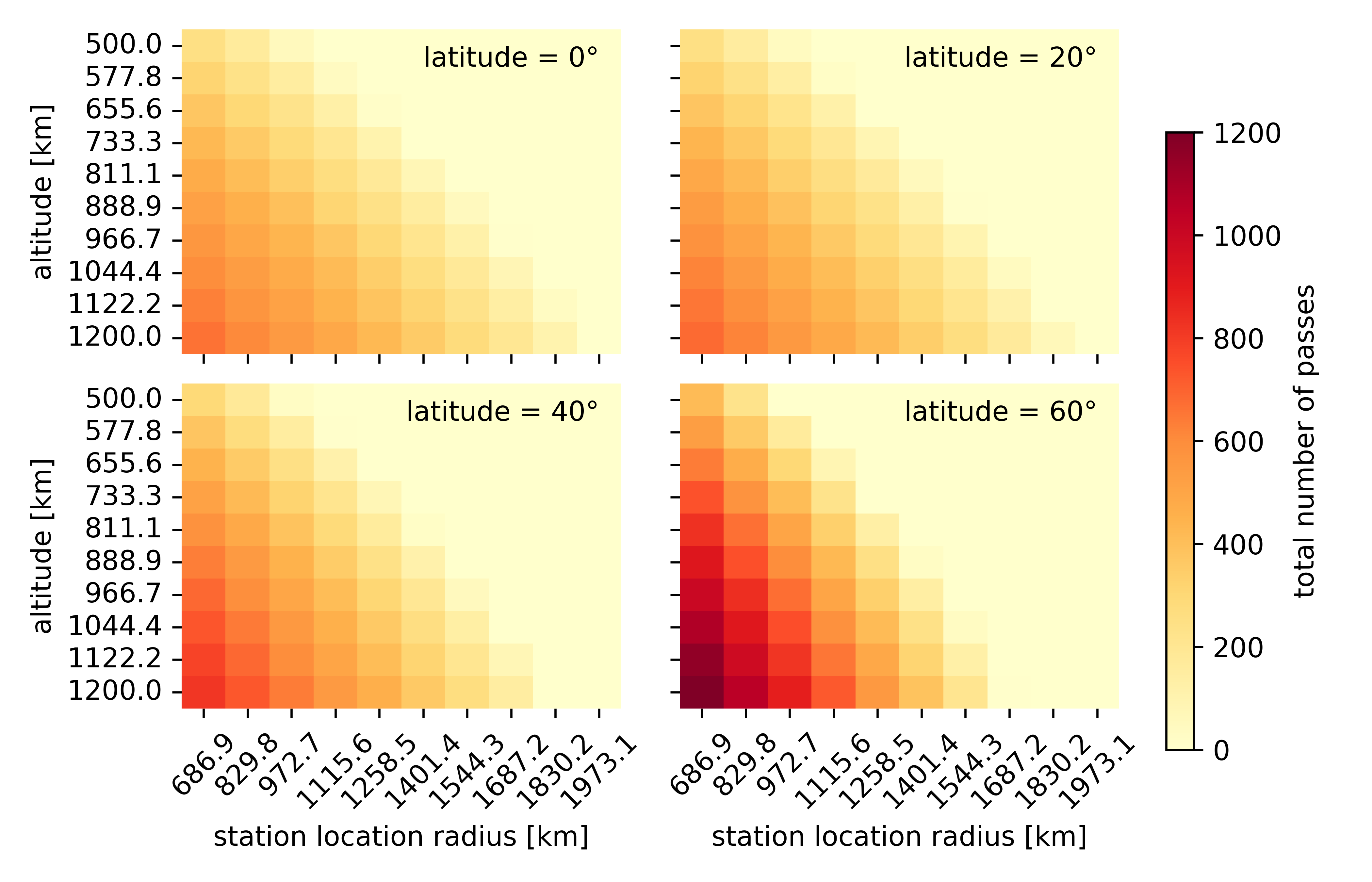}
  }\\
\end{tabular}
\caption{Heat maps showing the total number of passes as a function of station location radius and altitude with each subplot corresponding to different latitudes for inclinations $53^o$ and $86^o$. Darker colour indicates higher number of passes. }
\label{fig:total passes}
\end{figure*}

\begin{figure*}
\centering
\begin{tabular}{c}
  \subcaptionbox{Inclination $53^o$\label{fig:triangulation error (a)}}{%
    \includegraphics[width=\textwidth]{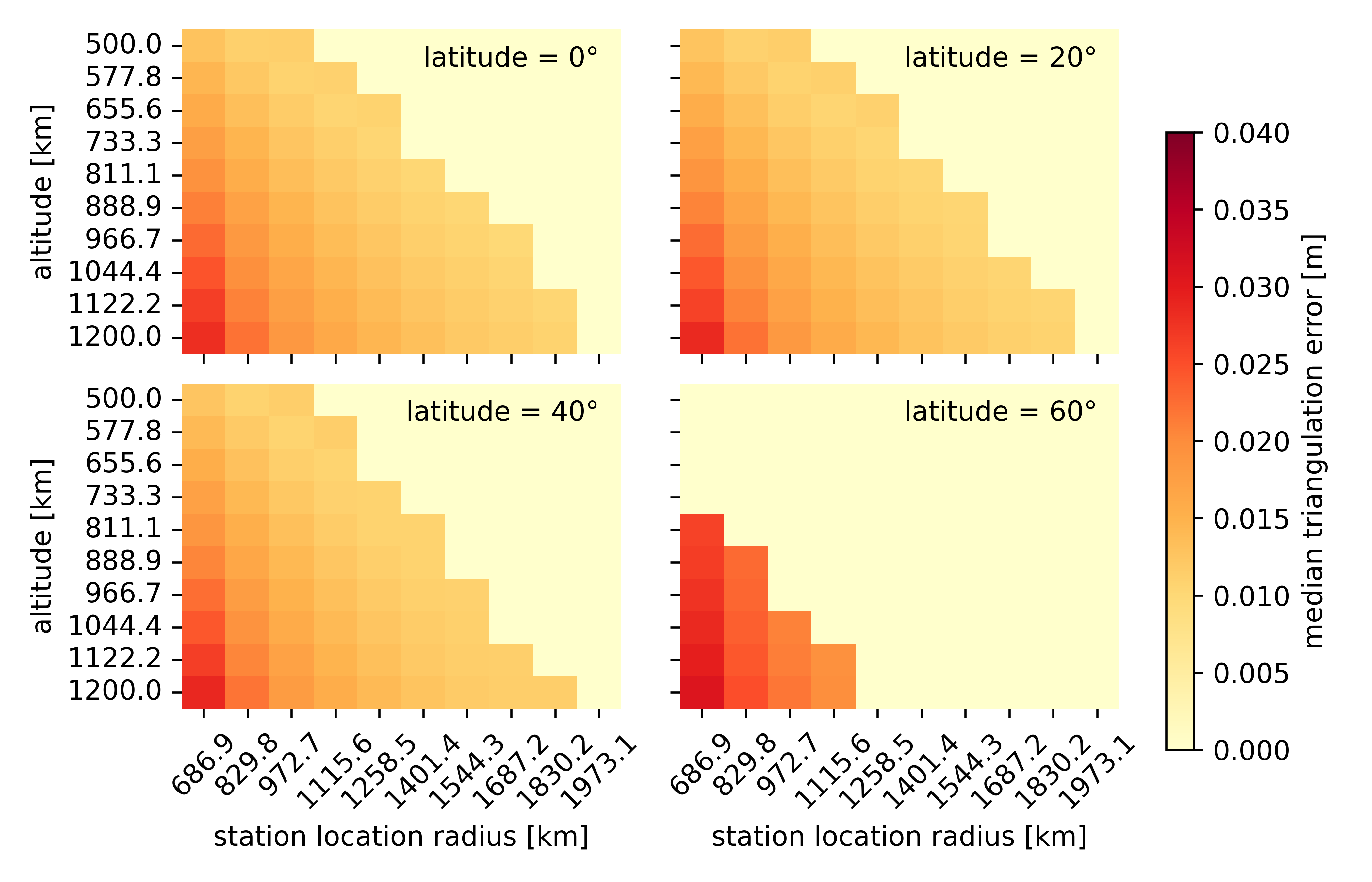}
  }\\[1ex]
  \subcaptionbox{Inclination $86^o$\label{fig:triangulation error (b)}}{%
    \includegraphics[width=\textwidth]{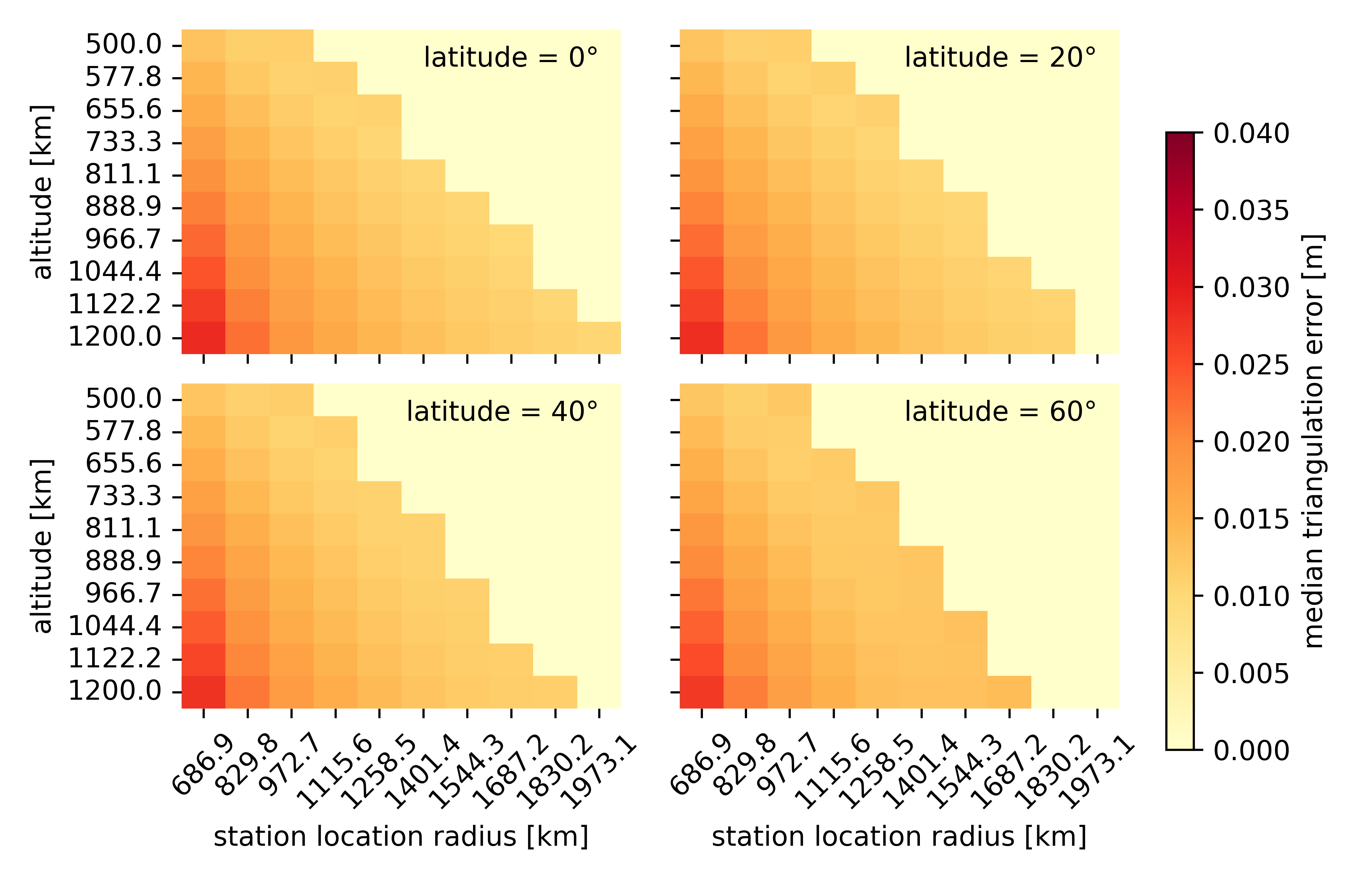}
  }\\
\end{tabular}
\caption{Heat maps showing the median triangulation error over a pass as a function of station location radius and altitude with each subplot corresponding to different latitudes for inclinations $53^o$ and $86^o$. A lower triangulation error indicates favourable satellite-to-ground station geometry.}
\label{fig:triangulation error}
\end{figure*}

\begin{figure*}
\centering
\begin{tabular}{c}
  \subcaptionbox{Inclination $53^o$\label{fig:pass duration (a)}}{%
    \includegraphics[width=\textwidth]{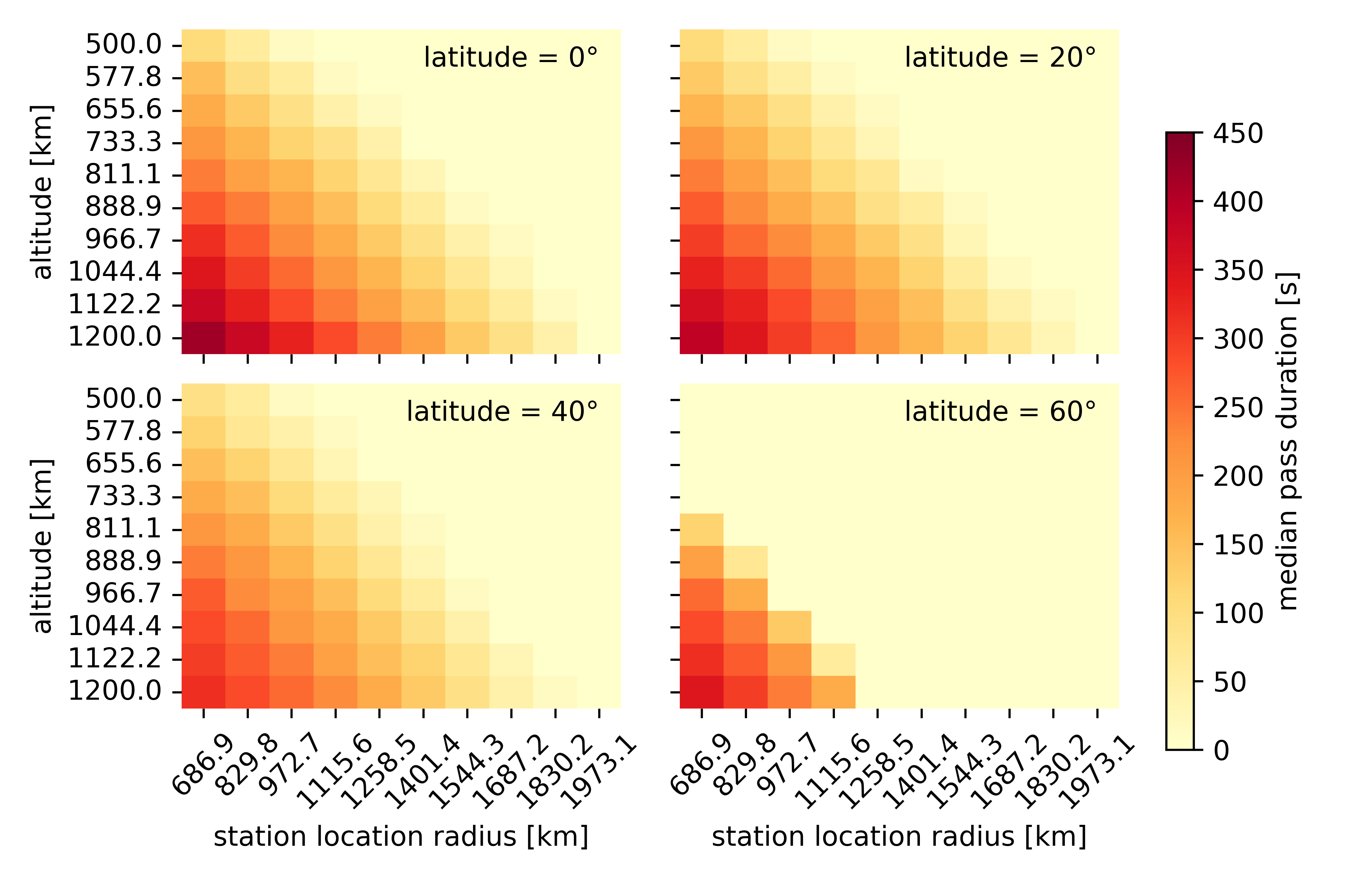}
  }\\[1ex]
  \subcaptionbox{Inclination $86^o$\label{fig:pass duration (b)}}{%
    \includegraphics[width=\textwidth]{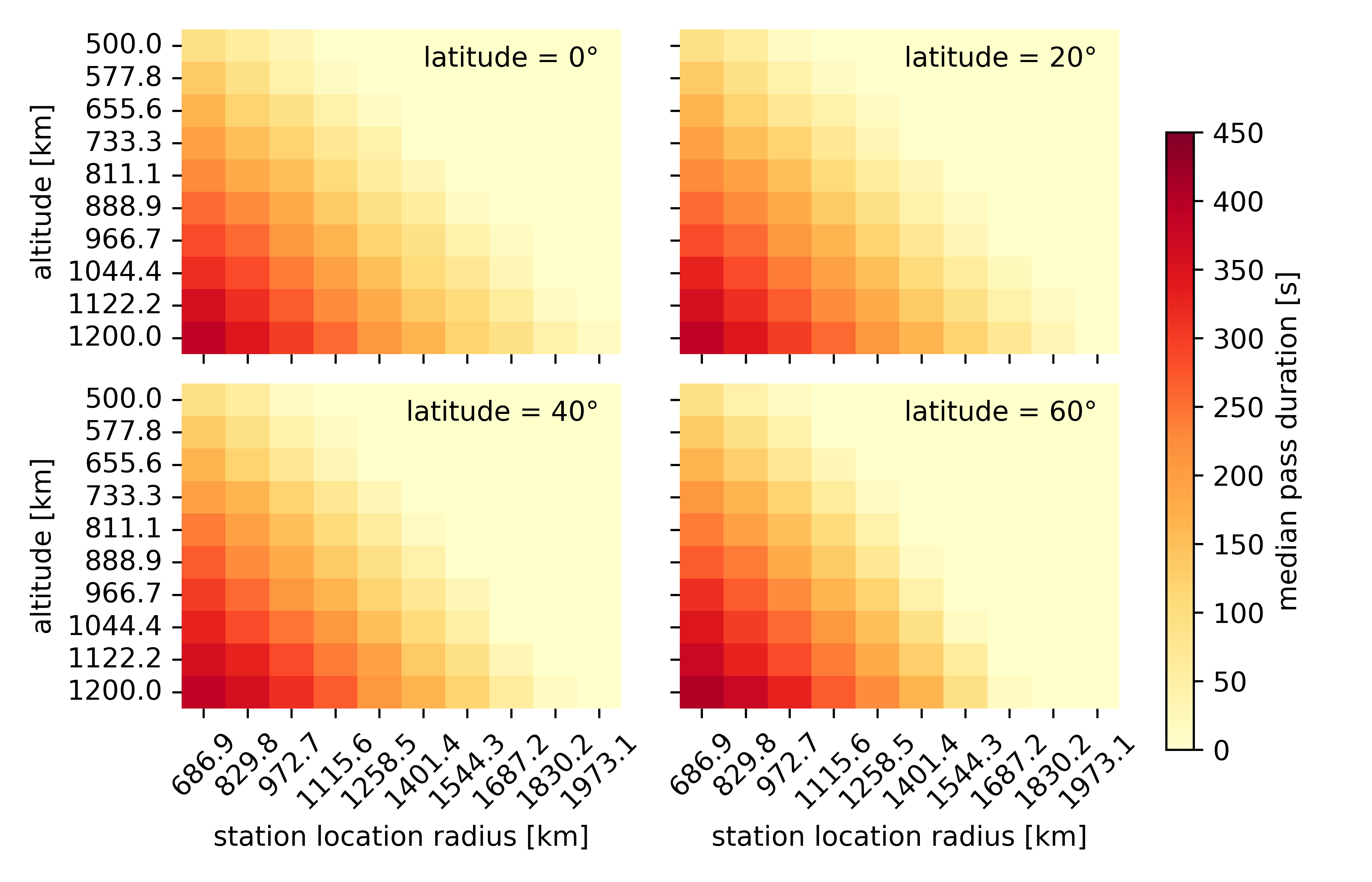}
  }\\
\end{tabular}
\caption{Heat maps showing the median pass duration as a function of station location radius and altitude, with each subplot corresponding to different latitudes for inclinations $53^o$ and $86^o$.}
\label{fig:pass duration}
\end{figure*}

A practical consideration for tri-static SLR-based methods is the frequency of high quality of observations that can be expected and how this relates to attitude determination. In this section, we show the metric that we use for determining the quality of a predicted pass over three ground stations. We then run a batch of simulations to see how many high quality passes are visible over tri-static SLR networks of various sizes for satellites across different regions of LEO.     

\subsection{How good is a given ground station pass?}
\label{sec: ground station placement passes}
Here, we provide a metric for how good a pass, or indeed section of a pass is, using the Fisher Information Matrix (FIM) to encapsulate the relationship between the ground station locations, the satellite locations, and the laser precision. Each ground station provides a line‐of‐sight measurement to the satellite, represented by a unit vector $\mathbf{u}_i$. Under the assumption of independent Gaussian measurement errors with variance $\sigma_i^2$, the contribution from the $i_\textrm{th}$ station to the FIM is given by

$${J}_i = \frac{1}{\sigma_i^2} \mathbf{u}_i \mathbf{u}_i^T.$$

Aggregating the contributions from all G ground stations, the total FIM becomes

$${J} = \sum_{i=1}^{G} \frac{1}{\sigma_i^2} \mathbf{u}_i \mathbf{u}_i^T.$$

The FIM encapsulates how sensitive the measurements are to changes in the state parameters. A more favorable geometric configuration, e.g. where the unit vectors $\mathbf{u}_i$ are well dispersed and nearly orthogonal, yields a well-conditioned ${J}$ and thus lower expected estimation errors.

According to the Cramér-Rao Lower Bound the covariance matrix ${C}$ of any unbiased estimator is bounded by the inverse of the FIM

$$ {C} \geq {J}^{-1}, $$

and for unbiased estimators the covariance can be approximated as

$$ {C} \approx {J}^{-1}. $$

This covariance matrix provides a quantitative measure of the expected error in the triangulation stage of the attitude determination process. In essence, it reflects how both the precision of individual measurements, through $\sigma_i^2$, and the spatial configuration of the ground stations, through $\mathbf{u}_i$, contribute to the overall uncertainty in the estimation of the satellite LRR locations in inertial space. Ergo, we use the square root of the trace of this covariance as a metric for how good a satellite pass over three ground station is. In simulation, we find heuristically that this metric is a strong predictor of the triangulation error and that this error can vary greatly over even a single pass as the relative satellite to station geometries changes.

\subsection{Ground station placement experiments}
\label{sec: ground station placement experiments}
It is important to demonstrate that a network of three ground stations can consistently obtain visible passes suitable for performing attitude determination, and that such a network can service satellites at different altitudes. To investigate this further, we performed ground station placement experiments to explore how different ground station configurations impact the number of passes and using the previously introduced measurement covariance metric to assess the pass quality. The desired outcome of this experiment is to show that across all LEO altitudes and inclinations, it is possible that a network of three SLR stations consistently yields a good number of passes, favourable pass durations, and minimal triangulation error. Additionally, the aim is to also identify an optimal baseline station configuration that can be used in future experiments, reducing the complexity of parameter selection. 

The experiments place three ground station in an equilateral triangle configuration, with a station at each vertex, of varying sizes and at four latitudes, $0^o$, $20^o$, $40^o$ and $60^o$. We then numerically integrate satellites at two inclinations, $53^o$ and $86^o$, forward for a period of one year and record the parameters of each of the passes over the network of stations.
For each inclination and latitude, we also explore the parameter space of satellite altitudes ranging from $500$~km to $1200$~ km. The final parameter we explore is the spacing of the ground stations in the equilateral triangle. We vary the radius of the circle centred on the centroid of the triangle intersecting all three ground stations, the station location radius, from $686.9$~km to $1973.1$~ km. This corresponds to a satellite at $300$~km altitude being visible at a $20^o$ elevation in the ground station topocentric reference frame up to a satellite at $1200$~km. We neglect weather conditions in these experiments but briefly mention results of some experiments that are inclusive of weather conditions.


The results were visualized using separate heat maps for the two different satellite inclinations and latitudes. Each heat map shows the station location radius on the x-axis and altitude on the y-axis, with the colour bar representing the total number of passes, median triangulation error and median pass duration. 

Figures~\ref{fig:total passes (a)}  and ~\ref{fig:total passes (b)}  show the heat map for the total number of passes with respect to altitude and station spacing for inclinations $53^o$ and $86^o$. Fig.~\ref{fig:total passes (a)} shows that the total number of passes increases at higher altitudes and smaller  station location radii, with latitudes $40^o$ and $60^o$ exceeding 1000 passes in a year. We see, unsurprisingly, that a station location radius that is too high results in a lack of available passes for attitude determination. This is less of a problem for satellites at higher altitudes but in order to cover the whole of LEO the data suggests that the station location radius should be kept at or below $830$~km for both inclinations. Also of note is that station networks at higher latitudes are preferable for high altitude satellite attitude determination whereas lower latitudes are preferable for lower altitudes.

Fig.~\ref{fig:triangulation error}  shows the median triangulation error with respect to altitude and station spacing for the same inclinations as before. It can be seen that for altitudes above $1044.4$~km  and station spacing of $686.9$~km, the median triangulation error is the highest and it decreases as the spacing increases. Therefore, a station network at $830$~km here also seems preferable for being able to service the whole of LEO.  The trend for both inclinations are similar except for the latitude $60^o$ scenario where the error is above $0.025$~m due to the satellite passes all being low on the horizon. The median triangulation error regions with zero should be considered as regions with no data as the total number of passes are zero in those regions. If selecting ground stations based on median triangulation error irrespective of inclination, it is better to avoid the latitude $60^o$ scenario. For all other latitudes, the station spacing is suitable as long as it is chosen to be greater than the smallest value observed in the heat map, which is approximately $686.9$~km.

With the total number of passes, it is also important to know the length of those passes to be useful for taking measurements. Fig.~\ref{fig:pass duration} shows the median pass duration with respect to altitude and station spacing and for the same inclinations as before. It can be seen from Fig.~\ref{fig:pass duration (a)}, the latitudes $0~^o$ and $20~^o$ scenario has a median pass duration above $300$~s for higher altitudes and smaller spacing. The major difference when compared to the total number of passes is that for latitude $40^o$, even though the total number of passes is high for higher altitudes and smaller station spacing, the pass duration is low. It is necessary to select lower spacing and higher altitude if the median pass duration needs to be high, although our method does not require a whole satellite rotation to be observer in order to function. 

The general trend is that the total number of passes and pass duration increase with altitude ($500$~km to $1200$~ km) and decrease with station spacing ($686.9$~km to $1973.1$~km). However, smaller station spacing also results in higher triangulation error. The selection of ground station placement should be such that there is good overall coverage which means a higher number of passes primarily but also reasonable pass duration. However, all this must be done while maintaining a low triangulation error. In summary, configurations of ground stations with station location radius of $829.8$~km with latitudes up to $40~^o$ provide a good number of passes with acceptable pass durations and low triangulation error. For these configurations, the total number of possible passes per year exceeds $400$, with a median pass duration greater than $200$~s, and with a median triangulation error below $0.020$~m. 

We also conducted experiments incorporating cloud cover data for specific locations and found that the latency, i.e. the time from an operator wanting attitude data to a pass being available, is seasonal, but that the maximum time an operator would have to wait for an attitude assessment is two weeks.

\section{Conclusions}
\label{sec: conclusions}

In this paper, we have presented a new numerical method for obtaining the attitude state of a satellite from satellite laser ranging data simultaneously gathered from three ground stations.  The method uses three on-board retroreflectors to enable a triangulation approach and a geometric model of the retroreflector placement on the satellite body to be used to ascertain the position of the three retroreflectors in inertial space. Importantly, this overcomes the labelling problem encountered by previous methods and therefore yields unique attitude solutions. We demonstrate through simulations that our approach can estimate the  spin rate and axis of a satellite with high precision, on the order of 0.1°/s  and 1°. We have been diligent to remove as many assumptions from the simulations as reasonably possible.

We performed a simulation campaign over a wide parameter space of potential retroreflector placements to show that the choice of layout is of paramount importance and that failure to respect these findings can reduce the effectiveness of the attitude determination process. To avoid this, we provide results that can be used by operators to correctly place their retroreflectors on their satellites ensuring that attitude determination can be performed effectively at the end of life. We suggest that retroreflectors are placed on a single panel such that their normal vectors are aligned. We also recommend that three, and only three, retroreflectors be visible at any one time. We find that the spacing required for this method is very practical for inclusion in a large number of satellites and that a triangle layout, with a retroreflector at each vertex, with an average side length of $0.7$~m is ample for reaching the highest levels of precision. 

We perform network location placement experiments to see if a tri-static network can gather enough data from passes to be viable and find that a network of stations placed in an equilateral configuration with stations at each vertex on a circle of radius $800$~km leads to a good design that obtains hundred of passes per satellite per year over the whole of LEO when weather information is excluded. However, when incorporating weather information for specific locations we find that the maximum period that an operator would have to wait between requesting an attitude determination to be performed and a suitable pass taking place is two weeks, but often only a couple of days.

We expect this technology to become practically applicable in the near future but given the low cost, complexity, and mass required, suggest that operators that wish to pre-empt this roll out and to prepare their satellites for disposal in the case of failure could do so through the inclusion of three retroreflectors in their satellite designs. We hope that the retroreflector geometric layout information herein is helpful in this direction. For particular mission requirements and placement constraints, we welcome interested parties to contact the authors and Lumi Space, for support in choosing the correct placement for their satellites to maximise attitude determination performance.

\end{document}